\documentclass[
prd,nofootinbib,preprintnumbers,preprint
]{revtex4}

\usepackage{axodraw4j}
\usepackage{pstricks}

\usepackage{amssymb}
\usepackage{amsmath}
\usepackage{amsfonts}
\usepackage{graphicx}
\usepackage{color}

\newcommand{\sn}{{\tilde{\nu}_{LSP}}}
\newcommand{\Ds}{\Delta m^2_{\odot}}
\newcommand{\Da}{\Delta m^2_{\rm{Atm}}}

\def\tb{\tan\beta}
\def\tbR{\tan\beta_R}

\def\gsim{\raise0.3ex\hbox{$\;>$\kern-0.75em\raise-1.1ex\hbox{$\sim\;$}}}
\def\lsim{\raise0.3ex\hbox{$\;<$\kern-0.75em\raise-1.1ex\hbox{$\sim\;$}}}

\begin{document}

\preprint{IFIC/12-67}  

\title{Sneutrino Dark Matter in Low-scale Seesaw Scenarios}

\author{Valentina De Romeri}\email{deromeri@ific.uv.es}
\affiliation{
AHEP Group, Instituto de F\'\i sica Corpuscular -- 
C.S.I.C./Universitat de Val\`encia Edificio de Institutos de Paterna, 
Apartado 22085, E--46071 Val\`encia, Spain}
\author{Martin Hirsch}\email{mahirsch@ific.uv.es}
\affiliation{
AHEP Group, Instituto de F\'\i sica Corpuscular -- 
C.S.I.C./Universitat de Val\`encia Edificio de Institutos de Paterna, 
Apartado 22085, E--46071 Val\`encia, Spain}

\begin{abstract}
We consider supersymmetric models in which sneutrinos are viable dark
matter candidates. These are either simple extensions of the Minimal
Supersymmetric Standard Model with additional singlet superfields,
such as the inverse or linear seesaw, or a model with an additional
U(1) group. All of these models can accomodate the observed small
neutrino masses and large mixings.  We investigate the properties of
sneutrinos as dark matter candidates in these scenarios.  We check for
phenomenological bounds, such as correct relic abundance, consistency
with direct detection cross section limits and laboratory constraints,
among others lepton flavour violating (LFV) charged lepton
decays.  While inverse and linear seesaw lead to different results for
LFV, both models have very similar dark matter phenomenology,
consistent with all experimental bounds. The extended gauge model
shows some additional and peculiar features due to the presence of an
extra gauge boson $Z'$ and an additional light Higgs. Specifically, we
point out that for sneutrino LSPs there is a strong constraint on the
mass of the $Z'$ due to the experimental bounds on the direct detecton
scattering cross section.

\end{abstract}

\maketitle

\section{Introduction}

Although the existence of dark matter (DM) is supported by a variety
of astrophysical data, its identity is unknown. Many particle physics
candidates have been proposed to explain the DM \cite{Bertone:2004pz}.
In supersymmetric extensions of the standard model (SM) there are the
lightest neutralino and the scalar neutrino, which could both provide
the correct relic density for the DM \cite{Jungman:1995df}.  The
neutralino as a DM candidate has been studied in literally hundreds of
publications, but also sneutrinos as candidates for the cold dark
matter have actually quite a long history
\cite{Ibanez:1983kw,Hagelin:1984wv,Freese:1985qw}. However, ordinary
left sneutrinos, i.e. the sneutrinos of the MSSM (Minimal
Supersymmetric Extension of the SM), have been ruled out
\cite{Falk:1994es} as the dominant component of the dark matter in the
galaxy a long time ago due to their large direct detection cross
section \cite{Goodman:1984dc}. This leaves only ``mixed'' sneutrinos,
i.e. sneutrinos which are partly singlets under the SM group, as good
DM candidates. Motivated by neutrino oscillation data
\cite{Tortola:2012te}, in this paper we study scalar neutrinos as DM
candidates in models with a low-scale seesaw mechanism, either 
 MSSM-like models with an inverse \cite{Mohapatra:1986bd} 
or the linear seesaw \cite{Akhmedov:1995vm,Akhmedov:1995ip} or 
models based on an $U(1)_{B-L}\times U(1)_R$ extension of the MSSM 
group \cite{Hirsch:2011hg,Hirsch:2012kv}.

Singlet sneutrinos as DM have been studied in the literature before.
Neutrino masses require that pure Dirac sneutrino must have tiny
Yukawa couplings. Unless the trilinear parameters are
huge, Dirac (right) sneutrinos are therefore never in thermal
equilibrium in the early universe \cite{Asaka:2005cn,Asaka:2006fs}.
\footnote{Unless Dirac neutrino masses are due to a tiny vev of a
  non-standard Higgs field \cite{Choi:2012ap}. In this case, Dirac
  sneutrinos could be the DM and even explain the much discussed claim
  for a tentative $130$ GeV $\gamma$ line in the FERMI data
  \cite{Weniger:2012tx}.}  However, they could still be non-thermal DM
produced in the decay of the NLSP (``next-to-lightest supersymmetric
particle'') \cite{Gopalakrishna:2006kr}. Also, trilinear terms are
usually thought to be proportional to the associated Yukawa couplings,
$T_{\nu} \propto Y_{\nu} A \sim {\cal O}(1)$ eV. Treating $T_{\nu}$ as
a free parameter of the order of ${\cal O}(100)$ GeV, Dirac
sneutrinos can be made good thermal DM candidates, as has been
discussed in \cite{Borzumati:2000mc,Thomas:2007bu,Dumont:2012ee}.
Very light mixed sneutrinos of this type have been studied in
\cite{Belanger:2010cd}. The LHC phenomenology of mixed Dirac sneutrino
DM was studied in \cite{Belanger:2011ny}. Alternatively to a large
A-term, Dirac sneutrinos could also be made thermal DM in models with
an extended gauge group \cite{Lee:2007mt,Belanger:2011rs}.

In the classical seesaw picture
\cite{Minkowski:1977sc,seesaw,Mohapatra:1979ia,Schechter:1980gr}
lepton number is broken at a very large energy scale, possibly close
to the unification scale.  In such a setup also the right sneutrinos
are very heavy and decouple; the sneutrinos remaining in the light 
spectrum are then very MSSM-like. One could, of course, simply put 
the scale of the seesaw low, say around the TeV scale. Yukawa couplings 
of the order of ${\cal O}(10^{-6})$ could fit neutrino data and the 
right sneutrinos are thermalized. In such an electro-weak scale seesaw  
right sneutrinos are overabundant unless (i) (again) a large trilinear 
parameter is assumed \cite{Arina:2007tm}; (ii) a new $U(1)$ group is 
introduced \cite{Bandyopadhyay:2011qm}; or (iii) sneutrinos have a 
large coupling to the NMSSM (``next-to-minimal Supersymmetric 
Standard Model'') singlet \cite{Cerdeno:2009dv,Cerdeno:2011qv}. 

However, the situation is different in extended seesaw schemes like
the inverse \cite{Mohapatra:1986bd} or the linear seesaw
\cite{Akhmedov:1995vm,Akhmedov:1995ip}. Here, additional singlets need
to be introduced, but the neutrino Dirac Yukawa couplings can take
essentially any value and it is the smallness of the inverse or linear
seesaw terms which ``explains'' the smallness of the observed neutrino
masses. In these setups the sneutrinos are highly mixed states.
Inverse seesaw sneutrino DM has been studied previously in
\cite{Arina:2008bb,Dev:2012ru}. Our work differs in several aspects
from these earlier papers. \cite{Arina:2008bb} calculated all masses
at tree-level and did not carry out a detailed fit to neutrino data,
while we use full 2-loop RGEs for the parameters, one-loop corrected
mass matrices and pay special attention to constraints from neutrino
masses. Also the paper \cite{Dev:2012ru} has some overlap with our work,
but concentrates more on collider phenomenology of the inverse seesaw
with sneutrino DM.

There are also some recent paper studying extended gauge groups. 
\cite{An:2011uq} studies inverse
seesaw in an $SU(2)_R$ extension of the MSSM. Also 
two papers based on sneutrinos in $U_{B-L}(1)\times U_Y(1)$ have been 
published recently. In \cite{Khalil:2011tb} an inverse seesaw is 
implemented in $U_{B-L}(1)\times U_Y(1)$. In \cite{Basso:2012gz}
sneutrino DM within the $U_{B-L}(1)\times U_Y(1)$ group was studied 
assuming a standard seesaw. 
However, none of the above papers has studied 
linear seesaw variants. Finally, we mention that part of the 
results discussed in this paper have been presented previously 
at conferences \cite{DeRomeri:2012pu}.

All our numerical calculations have been done using {\tt SPheno}
\cite{Porod:2003um,Porod:2011nf}, for which the necessary subroutines
were generated using the package {\tt SARAH}
\cite{Staub:2008uz,Staub:2009bi,Staub:2010jh}.  We have written the
SARAH input files for the inverse and the linear seesaw, while for the
$U(1)_{B-L}\times U(1)_R$ model we used the SARAH input files from
\cite{Hirsch:2012kv}. The calculation of the relic density of the LSP
is then done with \texttt{MicrOmegas} \cite{Belanger:2006is} version
\texttt{2.4.5} based on the \texttt{CalcHep} \cite{Belanger:2010st}
output of \texttt{SARAH}. To perform the scans we used a Mathematica
package (\texttt{SSP}) \cite{Staub:2011dp}.

The rest of this paper is organized as follows. In the next 
section we first recall the main features of the supersymmetric 
inverse and linear seesaws, before discussing briefly the 
minimal $U(1)_{B-L}\times U(1)_R$ extension of the standard 
model. In section (\ref{sec:cnstr}) we discuss phenomenological 
constraints on the parameter space of the different setups. 
In section (\ref{sec:snudm}) we then calculate the relic density 
and direct detection cross section. 
We conclude in section (\ref{sec:cncl}).

\section{Setup: Low scale seesaws and extended gauge groups}
\label{sect:setup}

In this section we briefly discuss the different setups, which we will
use in the numerical sections of the paper. We first discuss
supersymmetric inverse and linear seesaw, before recalling the main
features of the minimal $U(1)_{B-L}\times U(1)_R$ extension of the
MSSM. The latter can be realized with either inverse or linear seesaw,
but has some interesting additional features which are not covered by
either the inverse or linear seesaw extensions of the MSSM.

\subsection{Inverse and linear seesaw}
\label{sect:invlinSS}

In both, the inverse and the linear seesaws the particle content of
the MSSM is extended by two types of singlet superfields, $\hat\nu^c$
and $\hat S$. The former is assigned a $L=+1$, while the latter has
formally $L=-1$.  The total superpotential can be written as
\begin{equation}\label{eq:Wtot}
W = W_{\rm MSSM} + W_{\nu^c}  + W_{\rm ISS}+ W_{\rm LSS}
\end{equation}
Here, $W_{\rm MSSM}$ is the usual MSSM superpotential
\begin{equation}\label{eq:WMSSM}
W_{\rm MSSM} = Y_u\,\hat{u}\,\hat{q}\,\hat{H}_u\,
             - Y_d \,\hat{d}\,\hat{q}\,\hat{H}_d\,
             - Y_e \,\hat{e}^c\,\hat{l}\,\hat{H}_d\,
             +\mu\,\hat{H}_u\,\hat{H}_d\ .
\end{equation}
Lepton number conserving terms for the new singlet fields 
$\hat\nu^c$ (``right-handed neutrino'') and $\hat{S}$ can 
be written as
\begin{equation}\label{eq:Wnuc}
W_{\nu^c} = Y_{\nu}\,\hat{\nu}^c\,\hat{l}\,\hat{H}_u\,
            + M_R\,\hat{\nu}^c\,\hat{S}\ .
\end{equation}
The first term generates Dirac neutrino masses, once the $H_u$
acquires a vacuum expectation value, while the second term is a mass
term for the new singlet fields. In the inverse seesaw lepton number
is violated by the term
\begin{equation}\label{eq:WISS}
W_{\rm ISS}= \frac{1}{2} \mu_S \,\hat{S}\,\hat{S}\ ,
\end{equation}
while in the linear seesaw case one writes lepton number violation 
as:
\begin{equation}\label{eq:WLSS}
W_{\rm LSS} = Y_{SL}\,\hat{S}\,\hat{l}\,\hat{H}_u\ .
\end{equation}
In both cases, it is usually assumed that the lepton number violating
terms are small
\cite{Mohapatra:1986bd,Akhmedov:1995vm,Akhmedov:1995ip}, see also the
discussion in section (\ref{sec:cnstr}). The neutrino mass matrix 
and the resulting constraints on the model parameters are discussed 
in section (\ref{sec:numass}). 

In supersymmetric models with lepton number violation, also the scalar
neutrinos must have a lepton number violating mass term
\cite{Hirsch:1997vz}. This term, ${\tilde m}^2_{M}$, is given by the
difference between the eigenvalues of the real and imaginary
components of the scalar neutrinos. It is therefore convenient to
separate the sneutrino mass matrix into CP-even and CP-odd blocks
\cite{Grossman:1997is}: \footnote{Separation into CP-even and CP-odd 
blocks requires CP-conservation, i.e. all parameters in the mass 
matrices below have to be real.}
\begin{eqnarray}
{\cal M}^2  =
\left(\begin{array}{cc} 
{\cal M}^2_+ & {\bf 0} \\
 {\bf 0} & {\cal M}^2_-
\end{array}\right).
\label{eq:Snu}
\end{eqnarray}
Mass matrices for the scalar neutrinos are different in the inverse
and linear seesaws. At the tree-level, in the inverse seesaw the
${\cal M}_\pm^2$ are given by: \footnote{We correct some misprints in
\cite{Arina:2008bb,Hirsch:2009ra}}
\begin{eqnarray}
{\cal M}_{\pm,ISS}^2  = 
\left(\begin{array}{ccc} 
m^2_L+ D^2 + (m_D^Tm_D)  & A_{LR}^T & m_D^T M_R \\
A_{LR} & m^2_{\nu^c}+(M_RM_R^{T})+(m_Dm_D^{T}) & 
       \pm M_R\mu_S + B_{M_R} \\
M_R^T m_D & \pm \mu_S M_R^{T} + B_{M_R}^T 
& m^2_S+ \mu_S^2+M_R^TM_R \pm B_{\mu_S}
\end{array}\right)
\label{eq:SnuMpmISS}
\end{eqnarray}
Here, $D^2=\frac{1}{2} m^2_Z \cos 2\beta$ are the MSSM D-terms,
$m_D=\frac{1}{\sqrt{2}}v_u Y_\nu$, $A_{LR}= T_{Y_{\nu}}v_u-\mu m_D
{\rm cotg} \beta$, $B_{M_R}$ is the soft bilinear term, $T_{Y_{\nu}}$
is the soft trilinear and $m_L^2$, $m_{\nu^c}^2$ and $m_S^2$ are the
scalar soft masses for the doublet and the singlets respectively. Only
${\mu_S}$ and the corresponding bilinear soft term $B_{\mu_S}$ violate
lepton number and only these two come with different signs in the CP-even 
and CP-odd mass matrices.

For the linear seesaw one finds 
\begin{eqnarray}
&{\cal M}_{\pm,LSS}^2& \\ \nonumber
& = &
\left(\begin{array}{ccc} 
m^2_L+ D^2 + (m_D^Tm_D) +  (M_L^TM_L) &  A_{LR}^T \pm M_L^T M_R^T  & 
      m_D^T M_R \pm A_{LS}^T \\ 
A_{LR} \pm M_R M_L  &   m^2_{\nu^c} + (M_RM_R^{T})+(m_Dm_D^{T}) 
    & \pm  m_D M_L^T + B_{M_R} \\ 
M_R^T m_D \pm A_{LS}
    & \pm M_L m_D^{T} + B_{M_R}^T 
   & m^2_S + M_LM_L^{T} + M^T_RM_R
\end{array}\right)
\label{eq:SnuMpmLSS}
\end{eqnarray}
with all definitions as in eq. (\ref{eq:SnuMpmISS}) and $M_L =
\frac{1}{\sqrt{2}} v_u Y_ {SL}$ and $A_{LS}= T_{Y_{SL}}v_u-\mu M_L
{\rm cotg} \beta$.  In these simple setups all other mass matrices are
as in the MSSM and, therefore, not discussed here.

\subsection{Minimal $SU(3)_c \times SU(2)_L \times U(1)_{B-L}\times U(1)_R$ 
extension of the MSSM}
\label{sect:3211}

In order to explain why neutrinos are so much lighter than all other
matter particles, we have considered in the previous section two
variants of the seesaw which can, in principle, be implemented at
virtually any mass scale. Such seesaw schemes are actually most easily
realized in a particular class of extensions of the MSSM with an
extended gauge group \cite{Malinsky:2005bi,Dev:2009aw,DeRomeri:2011ie}
based on the $SO(10)$ breaking chains
\begin{eqnarray}
SO(10) & \to & SU(3)_c\times SU(2)_L\times SU(2)_R \times U(1)_{B-L}
 \to SU(3)_c\times SU(2)_L \times U(1)_Y \label{eq:onestep} \\ 
SO(10) & \to & SU(3)_c\times SU(2)_L\times SU(2)_R \times U(1)_{B-L} 
\label{eq:twostep}  \\ \nonumber
  &\to & SU(3)_c\times SU(2)_L\times U(1)_R \times U(1)_{B-L}
 \to SU(3)_c\times SU(2)_L \times U(1)_Y 
\end{eqnarray}
A MSSM-like gauge unification is in this case perfectly viable, and
compatible with a $U(1)_R \times U(1)_{B-L}$ stage stretching down to
TeV.  We will follow eq. (\ref{eq:twostep}), since this variant can be
realized with the minimal number of additional superfields with
respect to the MSSM particle content. This model
\cite{Malinsky:2005bi,DeRomeri:2011ie}, which we will call the minimal
$U(1)_{B-L}\times U(1)_R$ extension (mBLR, for short) has been studied
in two recent papers \cite{Hirsch:2011hg,Hirsch:2012kv}. We will
follow the notation of \cite{Hirsch:2012kv} quite closely.

\begin{table}
\begin{center} 
\begin{tabular}{|c|c|c|c|c|} 
\hline \hline 
\mbox{}\;\;\;\;\;\mbox{}& \; Superfield\;  & \; 
$SU(3)_c\times SU(2)_L\times U(1)_R\times U(1)_{B-L}$\;& \; Generations \; \\ 
\hline 
& \(\hat{Q}\)  
& \(({\bf 3},{\bf 2},0,+\frac{1}{6}) \) & 3 \\ 
&\(\hat{d^c}\) & 
\(({\bf \overline{3}},{\bf 1},+\frac{1}{2},-\frac{1}{6}) \)& 3 \\ 
&\(\hat{u^c}\) & 
\(({\bf \overline{3}},{\bf 1},-\frac{1}{2},-\frac{1}{6}) \)& 3 \\ 
&\(\hat{L}\)  & \(({\bf 1},{\bf 2},0,-\frac{1}{2}) \) & 3 \\ 
&\(\hat{e^c}\) & \(({\bf 1},{\bf 1},+\frac{1}{2},+\frac{1}{2}) \) & 3 \\ 
&\(\hat{\nu^c}\) & \(({\bf 1},{\bf 1},-\frac{1}{2},+\frac{1}{2}) \) & 3 \\ 
&\(\hat S\)& \(({\bf 1},{\bf 1},0,0) \) & 3 \\ 
\hline
&\(\hat{H}_u\)  
& \(({\bf 1},{\bf 2},+\frac{1}{2},0) \) & 1 \\ 
&\(\hat{H}_d\)  & \(({\bf 1},{\bf 2},-\frac{1}{2},0) \) & 1 \\ 
&\(\hat{\chi}_R\)  & \(({\bf 1},{\bf 1},+\frac{1}{2},-\frac{1}{2}) \) & 1 \\ 
&\(\hat{\bar{\chi}}_R\)  & \(({\bf 1},{\bf 1},-\frac{1}{2},+\frac{1}{2})\) 
& 1 \\ 
\hline \hline
\end{tabular} 
\end{center} 
\caption{\label{tab:fc}The Matter and Higgs sector field content of
the $U(1)_{R}\times U(1)_{B-L}$ model. Generation indices
have been suppressed. The $\hat S$ superfields are included to
generate neutrino masses via the inverse seesaw mechanism. Under
matter parity, the matter fields are odd while the Higgses are
even.}
\end{table}

 The particle content of the mBLR model is given in table
(\ref{tab:fc}). In this setup, the presence of $\hat\nu^c$ is required
for anomaly cancellation.  Breaking the $SU(2)_L \times
U(1)_{B-L}\times U(1)_R$ to $U(1)_{Q}$ requires additional Higgs
fields. The vev of the fields ${\chi}_R$ and ${\bar{\chi}}_R$ break
$U(1)_{B-L}\times U(1)_R$, while the vevs of $H_u$ and $H_d$ break
$SU(2)_L$ and $U(1)_Y$. Note that since $H_u$ and $H_d$ are charged
also under $U(1)_R$, in the mBLR new D-terms are generated in the mass
matrix for the scalars. These additional contributions with respect to
the MSSM allow to have a larger mass for the lightest MSSM-like
CP-even mass eigenstates and makes it possible to have a $m_{h^0}
\simeq 125$ GeV without constraints on the supersymmetric particle
spectrum \cite{Hirsch:2011hg,Hirsch:2012kv}.

Assuming matter parity \cite{Hirsch:2012kv}, apart from the MSSM 
superpotential the model also has the terms
\begin{eqnarray}
W_{S} & = & 
  Y_{\nu}\hat{\nu^c}\hat{L}\hat{H}_u
     +Y_s\hat{\nu^c}\hat{\chi}_R \hat{S} 
- \mu_{R}\hat{\bar{\chi}}_R\hat{\chi}_R
+ \mu_S \hat{S} \hat{S}.
\label{eq:superpot}
\end{eqnarray} 
The 2nd term generates $M_R= \frac{1}{\sqrt{2}} Y_{s}v_{\chi_R}$ while
the last term generates the inverse seesaw discussed above.  The
model can, in principle, also be written with a linear seesaw included
\cite{Malinsky:2005bi}. Note, that the model assigns lepton number
necessarily in a different way then discussed in the last subsection,
since here $B-L$ is gauged.  Thus, $B-L$ is broken by the vevs
of ${\chi}_R$ and ${\bar{\chi}}_R$. However, neutrino masses are
generated in exactly the same way as in the simpler inverse seesaw
model, discussed in the previous subsection.

It is useful to reparametrize the vevs in a notation similar 
to the MSSM, i.e.:
\begin{eqnarray}\label{eq:deftbs}
 v_R^2 = v_{\chi_R}^2 + v_{\bar{\chi}_R}^2 ~, ~~
 v^2 = v_d^2 + v_u^2 \\ \nonumber
 \tbR = \frac{v_{\chi_R}}{v_{\bar{\chi}_R}} ~, ~~
 \tb = \frac{v_u}{v_d}.
\end{eqnarray}
The mass of the new $Z'$-boson is approximately given by 
\cite{Hirsch:2012kv} 
\begin{equation}\label{eq:mZp}
 m_{Z'}^2 = \frac{g_R^4 v^2}{4 (g_{BL}^2 + g_R^2)} 
          + \frac{1}{4} (g_{BL}^2 + g_R^2) v_R^2\,. 
\end{equation}
Thus, $v_R$ must be larger than approximately $v_R\gsim 5$ TeV, see 
also next section. 

Mass matrices for all sfermions for this model can be found 
in \cite{Hirsch:2012kv}. For us the sneutrino mass matrix is 
most important. In the mBLR model it is given by the expression for
the inverse seesaw, with exception of $M_R= \frac{1}{\sqrt{2}} Y_{s}
v_{\chi_R}$ and new D-term contributions:
\begin{eqnarray}\label{eq:Dterms}
D^2_{LL} =  & \frac{1}{8}&  \left(
 (g_{BL}^2 + g_{BLR}^2 - g_{BL} g_{RBL}) v_R^2 \cos(2\beta_R)+
 (g_L^2 + g_R^2 + g_{BL} g_{RBL}) v^2 \cos 2\beta 
 \right) 
 \nonumber \\
D^2_{RR} = &  \frac{1}{8}& \Big(
 (g_{BL}^2 + g_R^2 + g_{BLR}^2  + g_{RBL}^2 -2 g_{BL} g_{RBL} - 2 g_R g_{BLR}) 
v_R^2 \cos(2\beta_R) 
 \nonumber \\
 &  & +  
 (g_R^2 + g_{RBL}^2 - g_{BL} g_{RBL} - g_R g_{BLR}) v^2 \cos 2\beta  \Big) 
\end{eqnarray}
Here, $D^2_{LL}$ replaces $D^2$ of the simpler models, while
$D^2_{RR}$ are the new D-terms in the ($\tilde\nu^c,\tilde\nu^c$) part
of the mass matrix. Due to the lower limit for the $Z'$ mass, see
eq. (\ref{eq:mZp}), and since the new D-terms in eq. (\ref{eq:Dterms})
can have either sign, the free parameter $\tan\beta_R$ is constrained
to be close to $\tbR \simeq 1$, otherwise either one of the sneutrinos
(or one of the charged sleptons) becomes tachyonic.

\section{Phenomenological constraints} 
\label{sec:cnstr}

In this section we discuss phenomenological constraints on the
parameter space of the different models. Below, we concentrate on
neutrino masses and lepton flavour violation. Other constraints 
on the model space come from SUSY searches at colliders, from 
$Z^0$ physics (LEP) and from the Higgs results of the LHC 
collaborations ATLAS \cite{ATLAS:2012gk} and CMS \cite{CMS:2012gu}.

``Heavy'' singlet neutrinos with mass below the $Z^0$ boson are
excluded by LEP experiments \cite{lepINHL}, which set limits on
$|U^{\nu}_{ij}|^2$ of the order of $10^{-3}$ to $10^{-5}$ for the
neutrino mass range from 3 GeV up to 80 GeV.  L3 has searched also for
heavy iso-singlet neutrinos decaying via $N \to l W$ and set limits which
range from $|U^{\nu}_{ij}|^2 \lsim 2 \times 10^{-3}$ for masses of 80 GeV to
$|U^{\nu}_{ij}|^2 \lsim 10^{-1}$ for masses of 200 GeV
\cite{Achard:2001qv}.  Most importantly, the invisible width of the
$Z^0$ boson \cite{Beringer:1900zz} puts an upper limit on the $3
\times 3$ sub-block $U^{\nu}_{ij}$, $i,j\le 3$, of the neutrino mixing
matrix: $\left|1 - \sum_{ij=1, i \le j}^3 \left| \sum_{k=1}^3
U^{\nu}_{ik} U^{\nu,*}_{jk} \right|^2 \right| < 0.009$ at the 3-$\sigma$
level even in the case that the new mostly singlet neutrinos are
heavier than the $Z^0$ boson \cite{Hirsch:2012kv}.
Finally, the $Z^0$ width rules out pure left sneutrinos lighter 
than approximately half of the $Z^0$ mass, but sneutrinos with 
suppressed coupling to the $Z^0$ below roughly $0.02-0.1$ with 
respect to the MSSM coupling and masses below $m_{\tilde\nu}\lsim 
40$ GeV are allowed.

In inverse seesaw models the Higgs can decay to heavy plus light
neutrino, if the heavy neutrino has a mass below the Higgs mass
\cite{Hirsch:2012kv,1207.2756}.  This limits the Yukawa couplings to
roughly below $|Y_{\nu}| \lsim 0.02$ for $M_R \lsim 120$ GeV from
measured data on the channel $h \to W W^* \to ll\nu\nu$ \cite{1207.2756}. 
For larger masses current Higgs searches provide essentially no 
constraint yet.

For the model with the extended gauge group searches for a 
new $Z'$ boson at the LHC provide important constraints. Both, 
CMS \cite{PAS-EXO12015} and ATLAS \cite{1209.2535} have searched 
for, but not observed any hints for, $Z'$'s within the context of 
different models. For the $U(1)_{B-L}\times U(1)_R$ model the 
limits are of the order of (roughly) $m_{Z'} \gsim (1.7-1.8)$ TeV.

SUSY searches at ATLAS \cite{ATLAS-CONF-2012-109} and CMS
\cite{CMS-PAS-SUS-11-016} provide lower limits on squark and gluino
masses. For example, in mSUGRA/CMSSM models with $\tan\beta=10$,
$A_0=0$ and $\mu>0$, squarks and gluinos of equal mass are excluded
for masses below 1500 GeV \cite{ATLAS-CONF-2012-109}. This limit
essentially rules out any value of $M_{1/2}$ below approximately
($600-700$) GeV for $m_0\lsim 1000$ GeV and $M_{1/2}$ below
($350-400$) GeV in the limit of large $m_0$ for pure CMSSM. Of course,
the observation of a new resonance with a mass around $125-126$ GeV
\cite{ATLAS:2012gk,CMS:2012gu}, if interpreted as the lightest Higgs
boson, provides important constraints on SUSY parameters as well.
However, these constraints are different for the different models we
study in this paper. We will discuss them therefore when we discuss
numerical scans in section (\ref{sec:snudm}).

\subsection{Neutrino masses}
\label{sec:numass}

\subsubsection{Inverse seesaw}

In the inverse seesaw the neutrino mass matrix can be written at
tree-level as
\begin{equation}
M^\nu=\left(
\begin{array}{ccc}
0  & m_{D}^T & 0  \\[2mm]
m_{D}& 0 & M_R\\[2mm]
0  & M_R^T & \mu_S 
\end{array} \right),
\label{eq:NuMassIS}
\end{equation}
The smallness of the observed neutrino masses is then usually 
explained as the hierarchy $\mu_S \ll m_D < M_R$. 

Following the notation of \cite{Santamaria:1993ah}, we can count the
number of {\rm physical} parameters of the model as $N_{\rm phys}=N_Y
- N_G + N_{G'}$. Here, $N_Y$ is the number of parameters in the Yukawa
matrices (or mass matrices), $G$ is the original symmetry group which
is broken into $G'$ by the presence of the Yukawas (or mass terms). In
table \ref{tab:dof} the counting for the inverse seesaw is summarized.

\begin{table}[h!]
\begin{center} 
\begin{tabular}{|c|c|c|c|} 
\hline \hline 
& Parameters & Moduli & Phases \\ 
 \hline \hline
$N_Y$ & $Y_e , Y_\nu , M_R$~~Dirac type & $3 \times n^2$ & $3 \times n^2$ \\
& & &  \\
$N_Y$ & $\mu_S$~~Majorana type & $\frac{n(n+1)}{2}$ & $\frac{n(n+1)}{2}$ \\
\hline
G & $U(n)_L \bigotimes U(n)_{\nu^c} \bigotimes U(n)_{e^c} \bigotimes U(n)_S$ &
  $4 \times \frac{n(n-1)}{2}$ & $4 \times \frac{n(n+1)}{2}$ \\
& & & \\
G' & no LF conservation & & \\
\hline
$N_{phys}$ & & 21 & 9\\
\hline \hline
\end{tabular}
\end{center} 
\caption{\label{tab:dof}
Parameter counting for MSSM with an inverse seesaw for three generations.}
\end{table}
After absorbing all unphysical parameters by field rotations, we find
a total of 30 real parameters, 21 moduli (12 masses and 9 mixing
angles) plus 9 phases. It is common practice to choose a basis in
which the charged lepton mass matrix (Yukawa: $Y_e$) is diagonal,
which fixes 3 parameters. The remaining parameters could be fixed by
going to a basis where $M_R$ is real and diagonal. In this case
$Y_\nu$ and $\mu_S$ are completely general, arbitrary matrices,
containing the remaining 24 free parameters. For fitting the neutrino
data, however, it is more useful to first rewrite the neutrino Yukawa
couplings using a generalization of the Casas-Ibarra parametrization
\cite{Casas:2001sr}.

Consider first the effective mass matrix of the light neutrinos for
the inverse seesaw.  It is given by
\begin{equation}
m_\nu^{eff} = m_D^T {M_R^T}^{-1} \mu_S M_R^{-1} m_D,
\label{eq:NuMaISS}
\end{equation}
We can rewrite $m_D$ as \cite{Forero:2011pc} 
\begin{equation}
\label{eq:ISdirac}
m_D = M_R^T V_{\mu}^T (\sqrt{\hat\mu_S})^{-1}R{\rm Diag}(\sqrt{m_{\nu_i}})U_{\nu}. 
\end{equation}
Here $U_{\nu}$ is the mixing matrix determined by the oscillation
experiments, in the basis where the charged lepton mass matrix is
diagonal, $m_{\nu_i}$ are the three light neutrino masses, $R$ is an
arbitrary real orthogonal $3\times 3$ matrix and $\hat\mu_S$ are the 
eigenvalues of the matrix $\mu_S$ with $V_{\mu}$ the matrix which 
diagonalizes $\mu_S$. 

Eqs (\ref{eq:NuMaISS}) and (\ref{eq:ISdirac}) allow to fit neutrino 
data in a straightforward way, if the tree-level contribution 
dominates, see below. Since one can always choose a basis where 
$M_R$ is diagonal, the flavour violation necessary to fit oscillation 
data resides in $m_D$ and in $\mu_S$. Particularly simple solutions 
are found, assuming either $\mu_S$ or $m_D$ are diagonal too.
For diagonal $\mu_S$, for example, one finds
\begin{equation}
\label{eq:ISDFit}
m_D = {\rm Diag}(\sqrt{\frac{m_{\nu_i}}{\mu_{S_i}}}M_{R_i})U_{\nu}.
\end{equation}

Oscillation experiments have determined the mass squared differences
and mixing angles of the active neutrinos with high precision, see for
example \cite{Tortola:2012te}. Recently also the last of the mixing
angles in the left-handed neutrino sector has been measured in two
reactor neutrino experiments, DAYA-BAY \cite{An:2012eh} and RENO
\cite{Ahn:2012nd}. With all these data, the situation can be
summarized as follows: The atmospheric neutrino mass squared
difference and angles are $\Delta(m^2_{\rm Atm}) = (2.31-2.74)\times
10^{-3}$ eV$^2$ (normal hierarchy) and $\sin^2\theta_{\rm Atm} =
0.36-0.68$, the solar parameters are $\Delta(m^2_{\odot}) =
(7.12-8.20)\times 10^{-5}$ eV$^2$ and $\sin^2\theta_{\odot} =
0.27-0.37$ and finally $\sin^2\theta_{13}=0.017-0.033$, all at 3
$\sigma$ c.l. \cite{Tortola:2012te}. Apart from the data on the
reactor angle, neutrino angles are still well-fitted by the
tribimaximal mixing ansatz \cite{Harrison:2002er}, which has
$\sin^2\theta_{\rm Atm} = 1/2$ and $\sin^2\theta_{\odot} = 1/3$.

The large atmospheric and solar angles require large off-diagonals 
in at least one of the two matrices $Y_{\nu}$ or $\mu_S$. For 
the case of strict normal hierarchy ($m_{\nu_1}\equiv 0$) and 
diagonal $\mu_S$, oscillation data can be well fitted to leading 
order in the small parameter $\sin\theta_{13}$ by
\begin{equation}\label{eq:TBMYnu}
Y_\nu = |Y_{\nu}| \left( \begin{array}{ccc}
 0 &   0 &  0 \\
 a &   a(1 - \frac{\sin\theta_{13}}{\sqrt{2}})& 
-a (1+ \frac{\sin\theta_{13}}{\sqrt{2}}) \\
 \sqrt{2}\sin\theta_{13} &  1 &  1 \\
\end{array}  \right) \,,
\end{equation}
with
\begin{equation}
a = \left( {\Ds}/{\Da} \right)^{\frac{1}{4}} \sim 0.4\;,
\end{equation}
where $|Y_{\nu}|$ can be easily calculated from $\mu_S$ and $M_R$.

\begin{figure}
\centering
\includegraphics[scale=0.6]{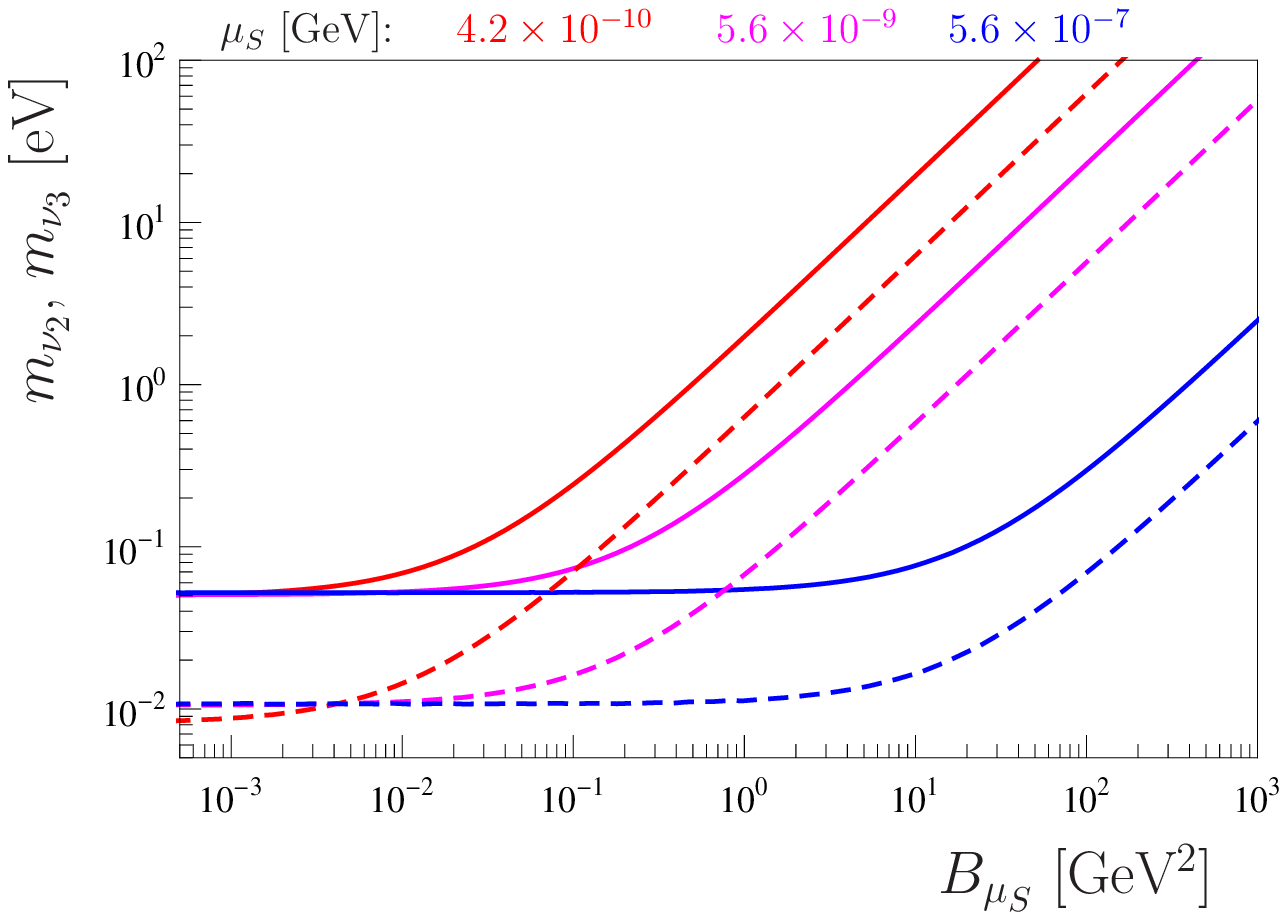} 
\includegraphics[scale=0.6]{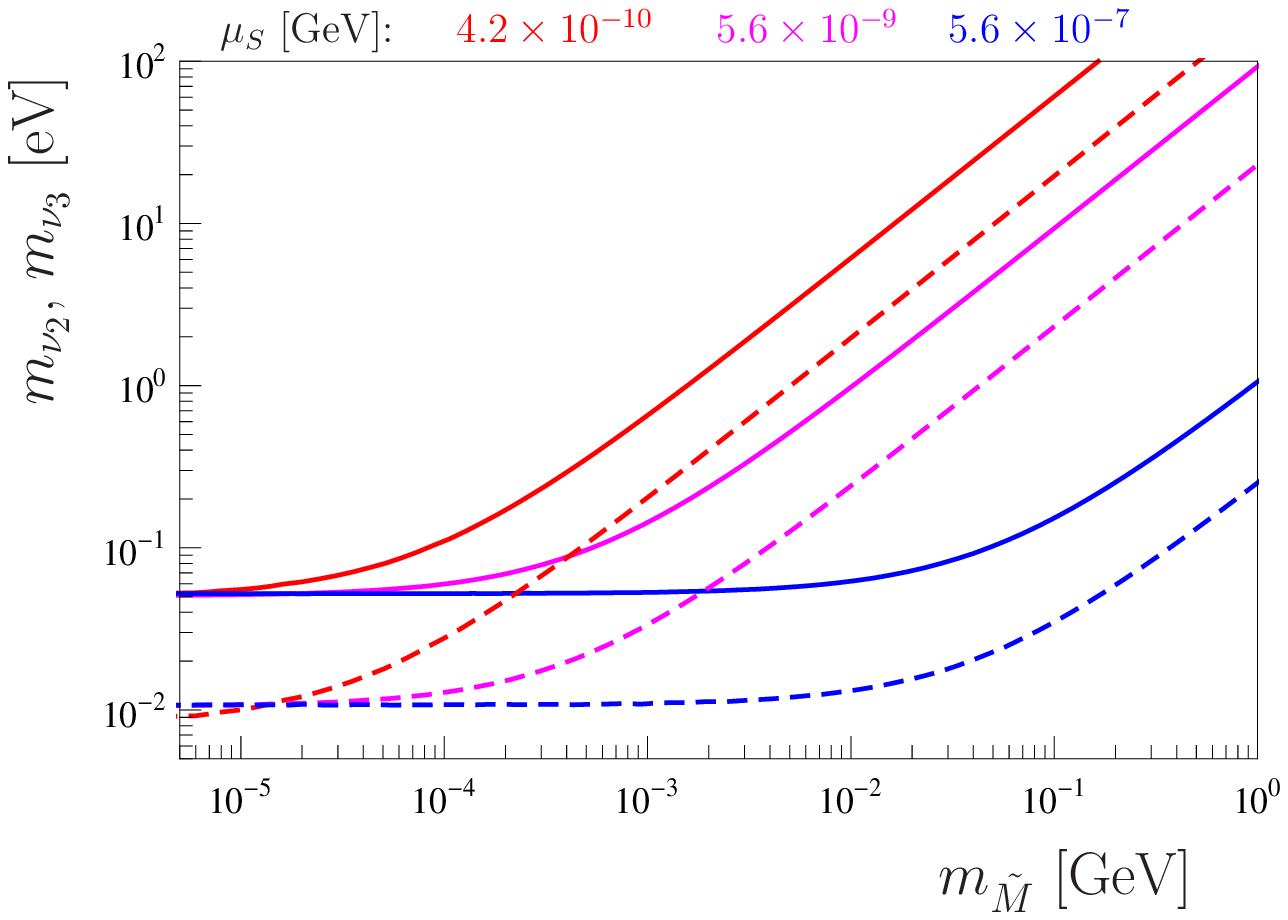}
\caption{\label{fig:split}Neutrino masses versus $B_{\mu_S}$ (left)
and versus $m_{\tilde M}$ (right) for one particular but arbitrary
parameter point (see text), for three different values of $\mu_S$.}
\end{figure}

The above discussion is valid at tree-level. In the inverse seesaw
neutrino masses also receive important corrections at the 1-loop
level, once $B_{\mu_S}$ becomes sizeable \cite{Hirsch:2009ra}.  An
example is shown in fig. (\ref{fig:split}). Here, we have chosen as an
example $m_0=100$, $M_{1/2}=1000$, $A_0=0$ (all in GeV)
$\tan\beta=10$, ${\rm sgn}(\mu)>0$ and $M_R=250$ GeV. For this plot
we assume $\mu_S$ and $B_{M_R}=3 \times 10^4$ GeV$^2$ to be diagonal
and degenerate. $Y_{\nu}$ is then fitted by eq.(\ref{eq:ISdirac}).
A smaller value of $\mu_S$ implies then a larger value for the entries 
in $Y_{\nu}$. 

In the left of fig. (\ref{fig:split}) we show $m_{\nu_2}$ and
$m_{\nu_3}$ as function of $B_{\mu_S}$, while the plot on the right
shows the same neutrino masses as a function of $m_{\tilde M}^2$, the
mass squared difference between the CP-even and CP-odd
sneutrinos. This splitting is proportional to $B_{\mu_S}$, while loop
neutrino masses are proportional to $Y_{\nu}^2 B_{\mu_S}$. To restrict
the neutrino mass to be smaller than the atmospheric scale than
results in an upper limit on $Y_{\nu}^2 B_{\mu_S}$. For $|\mu_S| \sim 5
\times 10^{-7}$ GeV, corresponding to the largest entries in $Y_{\nu}$
to be of order ${\cal O}(10^{-2})$, the splitting can be as large as
${\cal O}(10^{-1})$ GeV. Note, however, that with typical mSugra-like
boundary conditions one expects naively that $B_{\mu_S} \simeq \mu_S
m_0 \sim 10^{-4}-10^{-7}$ GeV$^2$. In this case splitting between 
the sneutrinos becomes negligible.

\subsubsection{Linear seesaw}
\label{subesct:linss}

\noindent
For the linear seesaw the neutrino mass matrix is given by
\begin{equation}
M^\nu=\left(
\begin{array}{ccc}
0  & m_{D}^T & M_L^T  \\[2mm]
m_{D}& 0 & M_R\\[2mm]
M_L  & M_R^T & 0 
\end{array} \right),
\label{eq:NuMassLS}
\end{equation}
with the effective neutrino mass matrix for the light neutrinos 
given as
\begin{equation}
\label{eq:linSmass}
m_\nu=m_D^T {M^T_R}^{-1}M_L+ M_L^TM_R^{-1}m_D.
\end{equation}
For the linear seesaw one finds for the CI parametrization 
\cite{Forero:2011pc} 
\begin{equation}
m_D = - M_R (M_L^T)^{-1} U_{\nu}^T \sqrt{m_{\nu_i}}A \sqrt{m_{\nu_i}} U_{\nu}
\label{eq:YvuLin}
\end{equation}
where $A$ has the following general form:
\begin{equation}
\left(
\begin{array}{ccc}
\frac{1}{2} & a & b\\
-a & \frac{1}{2}& c\\
-b & -c  & \frac{1}{2}
\end{array}
\right),
\label{eq:A}
\end{equation}
with $a,b,c$ real, but arbitrary numbers. The parameter counting 
for the linear seesaw is given in table (\ref{tab:dofLin}).
We have in total 36 real parameters, 24 moduli (12 masses and 12
mixing angles) plus 12 phases. Fits to neutrino data can be easily 
done using eqs (\ref{eq:linSmass}) and (\ref{eq:YvuLin}).

For example, for strict normal hierarchy, degenerate $M_R$ and diagonal 
and degenerate $Y_{SL}$ one finds to leading order in $\theta_{13}$
\begin{eqnarray}\label{eq:TBMYnuLinSS}
Y_\nu = |Y_{\nu}| 
\Large\{ &- & \frac{m_{\rm Atm}}{2}
\left( \begin{array}{ccc}
0 & \frac{\sin\theta_{13}}{\sqrt{2}} & \frac{\sin\theta_{13}}{\sqrt{2}}\\
\frac{\sin\theta_{13}}{\sqrt{2}} & \frac{1}{2} & \frac{1}{2} \\
\frac{\sin\theta_{13}}{\sqrt{2}} & \frac{1}{2} & \frac{1}{2} \\
\end{array}  \right) \\ \nonumber
&+& \frac{m_{\odot}}{3}\times \Big(
\left( \begin{array}{ccc}
 -1 & -1 & 1 \\
 -1 & -1 & 1 \\
  1 &  1 & -1 \\
\end{array}  \right) + 
\sqrt{2}\sin\theta_{13}\times 
\left( \begin{array}{ccc}
 0 & \frac{1}{2} & \frac{1}{2} \\
\frac{1}{2} & 1 & 0 \\
\frac{1}{2} & 0 & -1 \\
\end{array}  \right) \Big)\Large\},
\end{eqnarray}
where again, the prefactor $|Y_{\nu}|$ can be calculated from 
$|Y_{SL}|$ and $M_R$. Note that the flavour structure of 
eq. (\ref{eq:TBMYnuLinSS}) differs significantly from 
eq. (\ref{eq:TBMYnu}) for the same choice of angles, see the 
discussion about lepton flavour violation in the next subsection. 

\begin{table}[h!]
\begin{center} 
\begin{tabular}{|c|c|c|c|} 
\hline \hline 
& Parameters & Moduli & Phases \\ 
 \hline \hline
$N_Y$ & $Y_e , Y_\nu , Y_{SL}, M_R$~~Dirac type & $4 \times n^2$ & 
      $4 \times n^2$ \\
\hline
G & $U(n)_L \bigotimes U(n)_{\nu^c} \bigotimes U(n)_{e^c} \bigotimes U(n)_S$ 
 &  $-4 \times \frac{n(n-1)}{2}$ & $-4 \times \frac{n(n+1)}{2}$ \\
& & & \\
G' & no LF conservation & & \\
\hline
$N_{phys}$ & & 24 & 12\\
\hline \hline
\end{tabular}
\end{center} 
\caption{\label{tab:dofLin}Parameter counting for the linear seesaw model 
for three generations.}
\end{table}

In case of the linear seesaw, loop contributions to the neutrino
masses from the splitting in the sneutrino sector is always negligible
for neutrino masses in the sub-eV range, assuming the trilinears to be
proportional to $T_x \propto Y_x A_0$. This can be understood as
follows: The difference in the eigenvalues of the CP-even and CP-odd
sector is entirely due to the different signs in the off-diagonals in
eq. (\ref{eq:SnuMpmLSS}). As can be easily shown, the maximum
difference in the eigenvalues is reached for $Y_{SL}\simeq Y_{\nu}$. 
However, eq.(\ref{eq:linSmass}) shows that the product $Y_{SL} Y_{\nu}$ 
is required to be small, due to the observed smallness of neutrino 
masses. Thus, the splitting in the sneutrino sector in case of 
linear seesaw is maximally of the order of $m_{\nu}m_{SUSY}$, 
i.e. ${\cal O}(10^{-9})$ GeV$^2$.

\subsection{Lepton flavour violation}

In any supersymmetric model, limits on lepton flavour violating decays
such as $\mu\to e \gamma$ provide an important constraint on the 
parameter space \cite{Borzumati:1986qx}. In models with a low scale 
seesaw especially important constraints come from 
$l_i \to 3 l_j$ \cite{Hirsch:2012ax} and from $\mu -e$ conversion 
in nuclei \cite{Abada:2012cq}.

\begin{figure}
\centering
\includegraphics[scale=0.6]{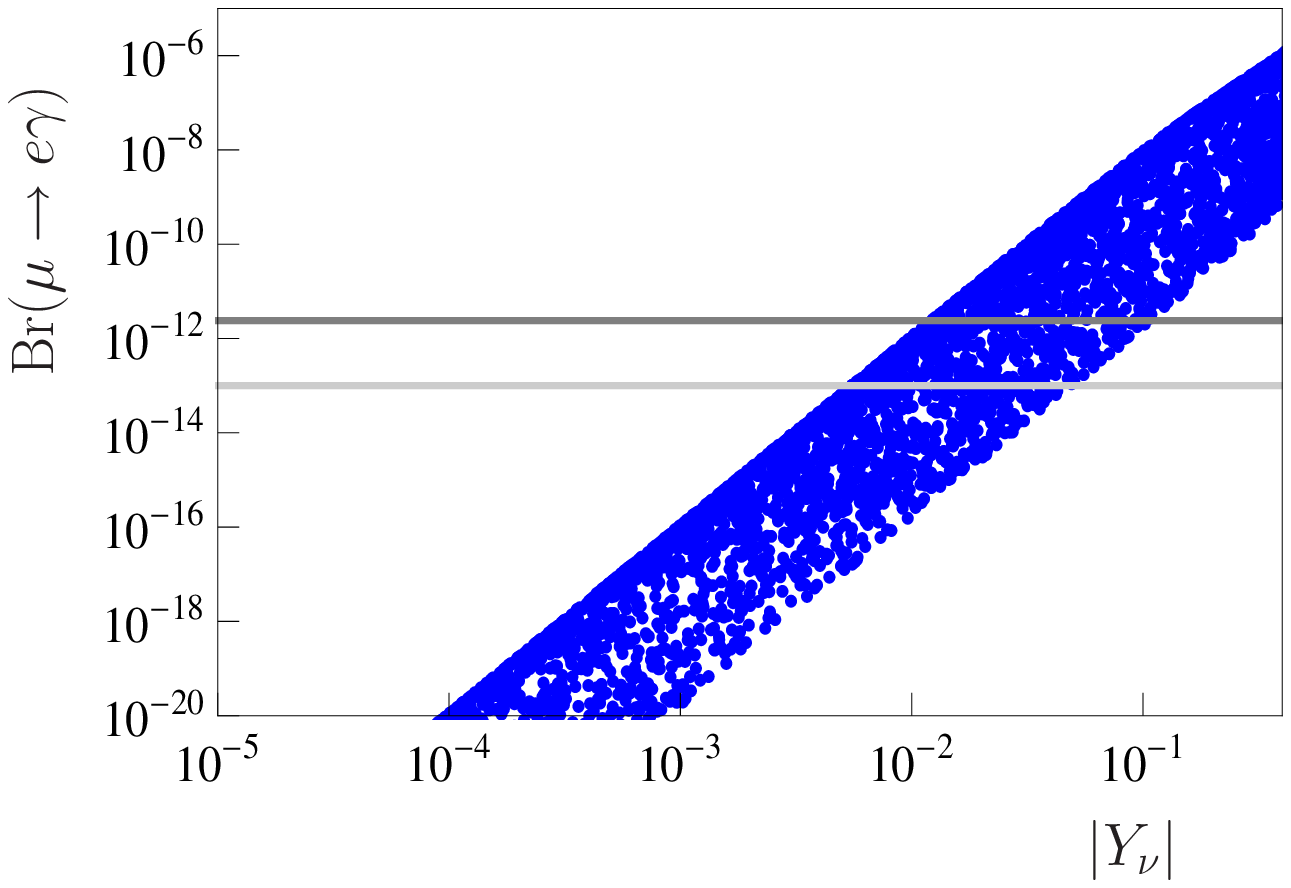}
\includegraphics[scale=0.6]{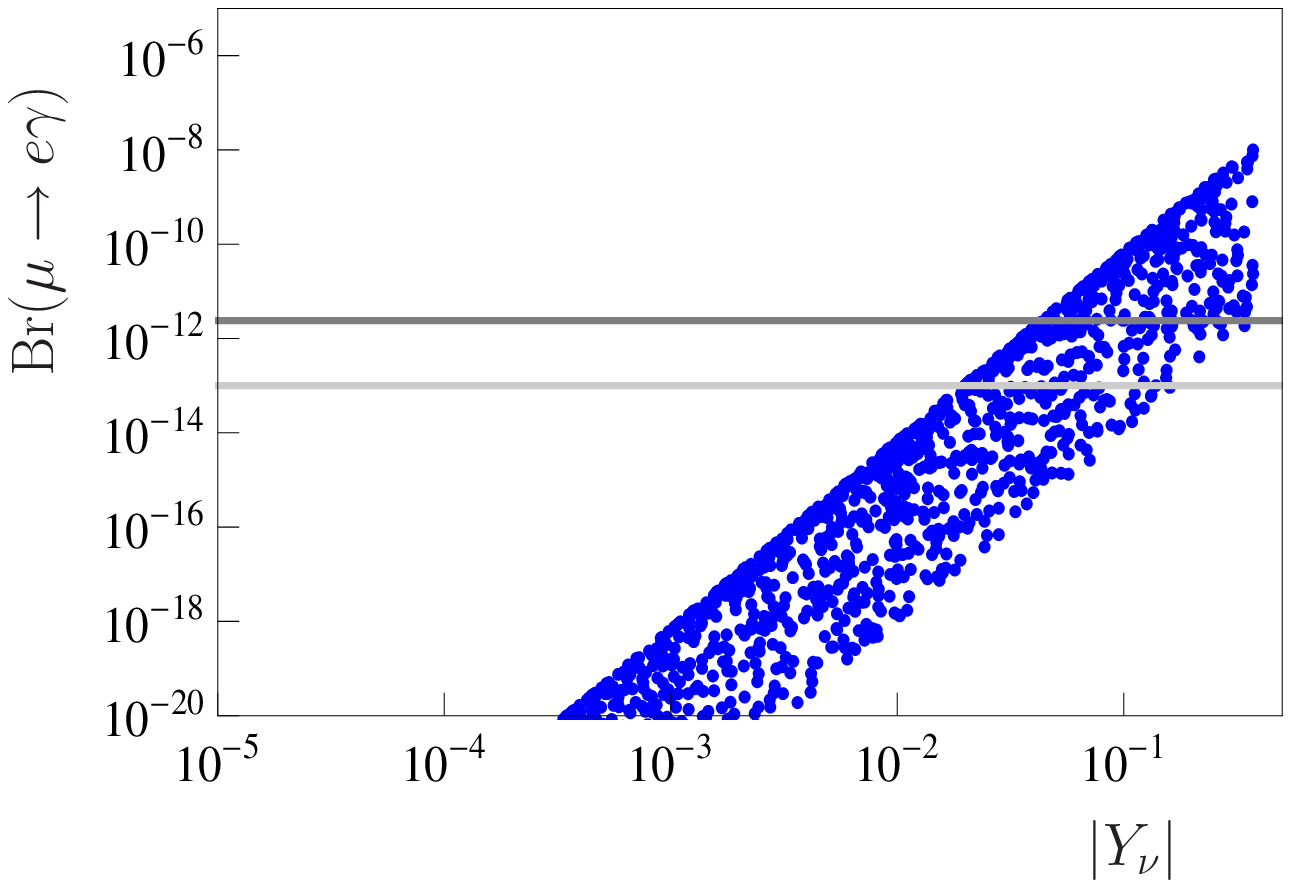}
\caption{Br($\mu\to e \gamma$) calculated in the inverse (left) and 
linear (right) seesaw. Here, all flavour has been put into the 
Yukawa $Y_{\nu}$, while neutrino angles have been fitted to their 
best fit point values \cite{Tortola:2012te}. A random scan over 
$m_0$ and $M_R$ in the interval [100,1000] GeV and $B_{M_R}=[10^3,10^6]$
GeV$^2$ has been performed for these points. Note that the axis are 
the same for inverse and linear seesaw for an easier comparison. 
Linear seesaw leads to smaller LFV than inverse seesaw for equal 
choice of neutrino angles. }
\label{fig:LFVInvSS}
\end{figure}

The fit to neutrino data requires non-trivial flavour violating
entries in at least one of the Yukawa- or mass matrices: $Y_{\nu}$ or
$Y_{SL}$ for linear and $Y_{\nu}$ or $\mu_S$ for inverse seesaw.  If
we assume that the LFV resides in $Y_{\nu}$, limits on the Yukawa
result as shown in fig. (\ref{fig:LFVInvSS}). In this figure we have
chosen $\mu_S$ (left) or $Y_{SL}$ (right) diagonal and neutrino angles
have been fitted to their best fit point values \cite{Tortola:2012te}
using $Y_{\nu}$.  A random scan over $m_0$ and $M_R$ in the interval
[100,1000] GeV and $B_{M_R}=[10^3,10^6]$ GeV$^2$ has been performed
for these points, fixing $\tan\beta=10$ and $M_{1/2}=2.5$ TeV. Upper
limits of the order of (few) $10^{-2}$ ($10^{-1}$) result for inverse
(linear) seesaw, despite the heavy SUSY spectrum (due to the
large value of $M_{1/2}$). Much stronger limits result for lighter spectra. 
Note that $l_i \to 3 l_j$ \cite{Hirsch:2012ax} and $\mu -e$ 
conversion in nuclei \cite{Abada:2012cq} can lead to even stronger 
limits. We will not repeat this exercise here.

Note also, as discussed in the next section, that the constraints from
relic density of sneutrinos lead to an approximate lower bound on the
absolute size of the Yukawa coupling $|Y_{\nu}|$.

\section{Sneutrino Dark Matter}
\label{sec:snudm}

In this section we discuss the relic abundance (RA) and the direct 
detection cross section (DD) of sneutrinos in the different models. 
We will first discuss the simpler case of the inverse/linear seesaw 
and then turn to the mBLR model.

In order to reduce the number of free parameters in our numerical 
scans, we calculate all spectra with CMSSM-like boundary conditions, 
i.e. at the GUT scale we choose ($m_0,M_{1/2},A_0,\tan\beta,{\rm sgn}(\mu)$), 
from which all soft parameters at the electro-weak scale are calculated 
using full 2-loop RGEs. Unless noted otherwise, we always assume that 
the trilinear soft parameters are related to the superpotential 
parameters in a ``mSugra''-like way: $T_{\alpha} \propto Y_{\alpha}A_0$ 
at $m_{GUT}$. 

In addition to the MSSM parameters, we have the neutrino Yukawa 
couplings $Y_{\nu}$ and several model specific parameters. These 
are $M_R$ and $B_{M_R}$ and, in case of the inverse (linear) seesaw 
$\mu_S$ and $B_{\mu_S}$ ($Y_{SL}$). While, in principle, all of these 
are matrices we use eq. (\ref{eq:ISdirac}) and (\ref{eq:YvuLin})) 
to fit neutrino data and usually assume all matrices are diagonal 
except one.

For the mBLR model we have the free parameters $Y_s$, $v_R$,
$\tan\beta_R$, $\mu_R$ and $m_{A_R}$. Recall, $M_R=Y_s v_R/\sqrt{2}$
and $m_{A_R}$ is the CP-odd scalar Higgs mass in the $\chi_R$
sector. Due to the constraints from LFV discussed above, we usually
put all LFV into either $\mu_S$ (inverse seesaw) or $Y_{SL}$ (linear
seesaw). This way we only have to check for the constraints from $Z^0$
and Higgs physics and lower limits on squarks and gluinos discussed in
section (\ref{sec:cnstr}).

\subsection{Inverse/Linear seesaw}
\label{sec:snudmISS}

\begin{figure}
\centering
\includegraphics[scale=0.6]{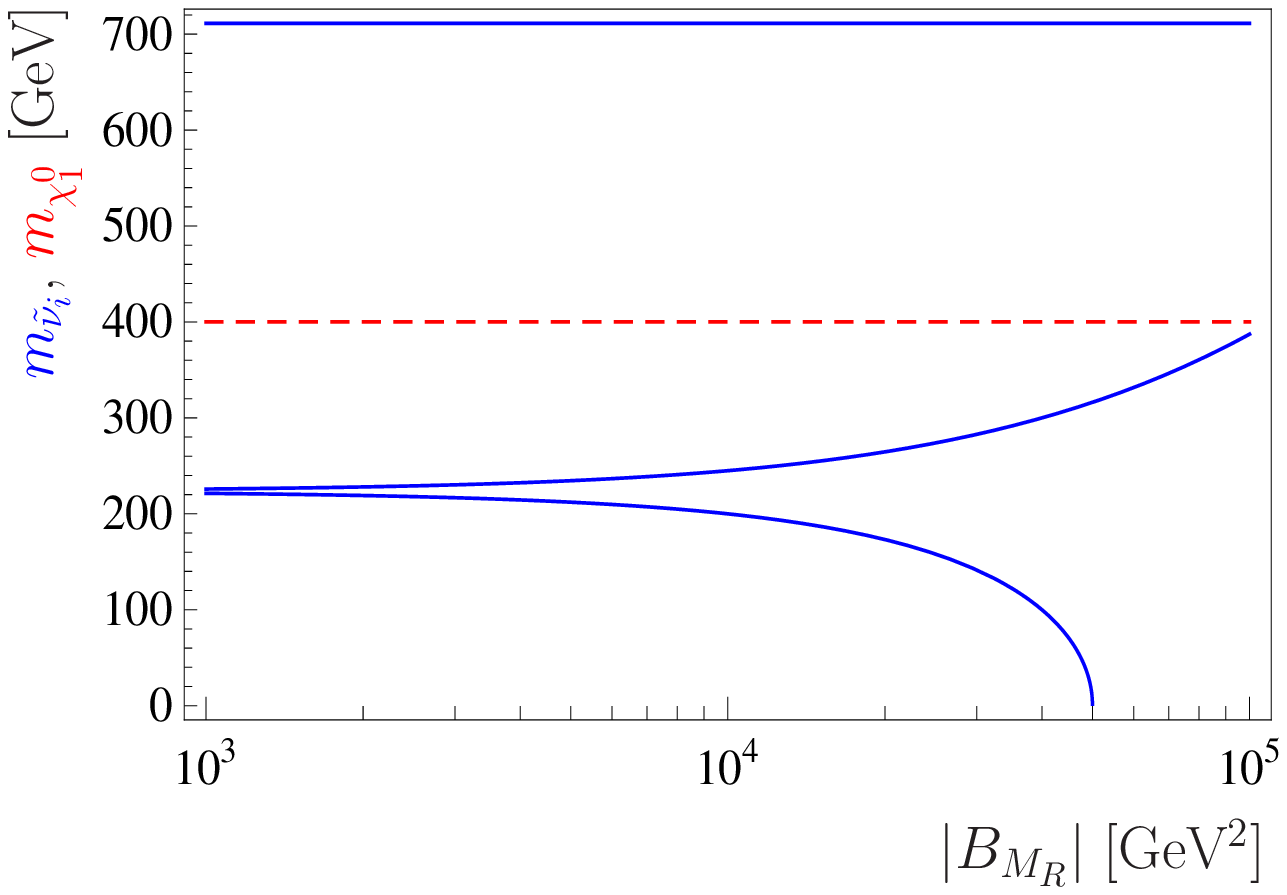}
\includegraphics[scale=0.6]{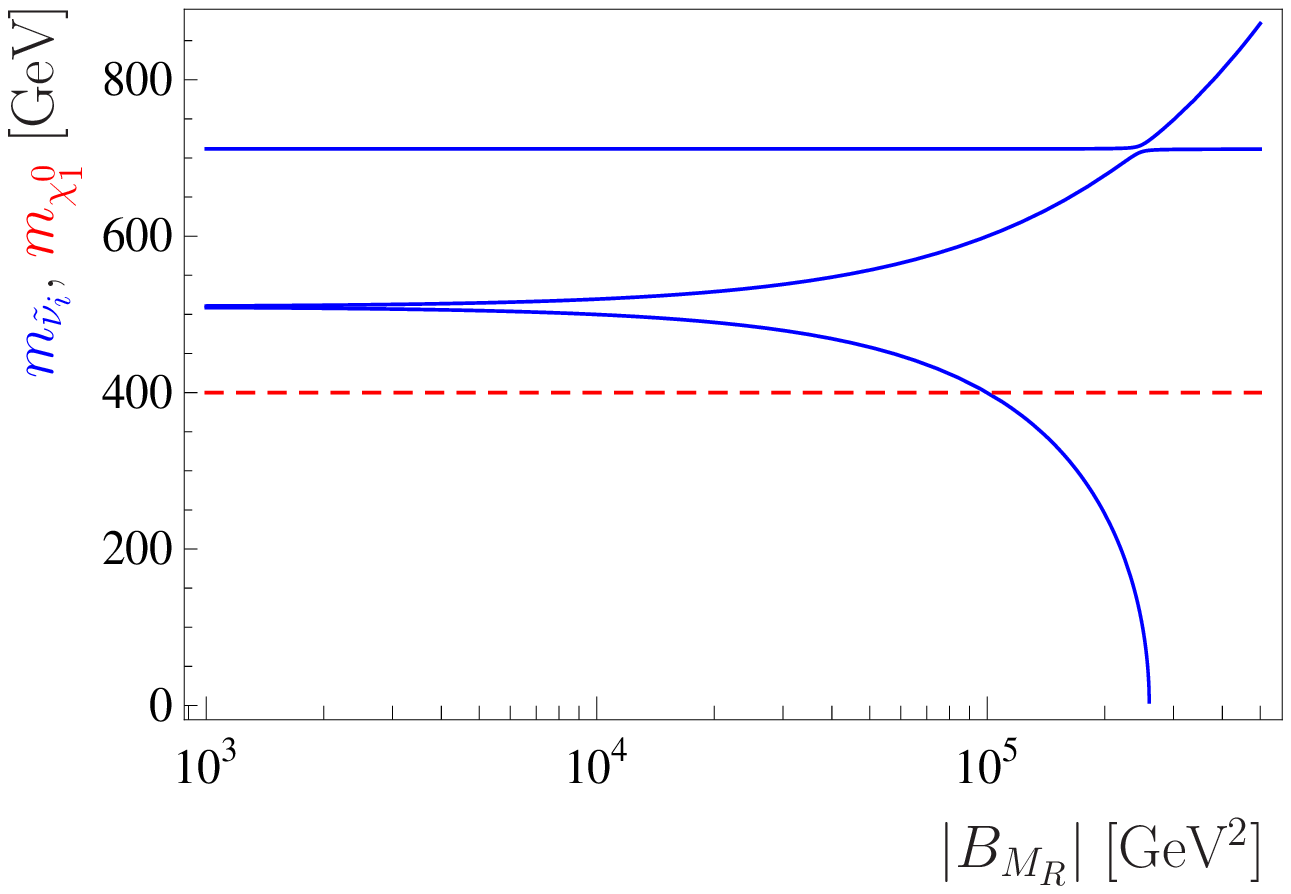}
\caption{\label{fig:Sneus}
Two examples of tree-level sneutrino masses calculated as
function of $B_{M_R}$ for two particular but arbitrary choices of
parameters: $m_0 = 100$, $M_{1/2} = 1000$, $A_0 = 0$ and $\mu = 800$
all in GeV and $|Y_{\nu}|=0.1$ and $\tan\beta = 10$. In addition
$M_R=200$ GeV (left) and $M_R=500$ GeV (right). For comparison also 
the lightest neutralino mass is shown. }
\end{figure}

Sneutrinos can be the LSP, practically independent from the actual 
choice of the CMSSM parameters. This can be easily understood from 
eqs (\ref{eq:SnuMpmISS}) and (\ref{eq:SnuMpmLSS}) and is demonstrated 
by two simple examples in fig. (\ref{fig:Sneus}).

In fig. (\ref{fig:Sneus}) we show two examples of tree-level sneutrino
masses calculated as function of $B_{M_R}$ for two particular but
arbitrary choices of parameters: $m_0 = 100$, $M_{1/2} = 1000$, $A_0 =
0$ and $\mu = 800$ all in GeV and $|Y_{\nu}|=0.1$ and $\tan\beta =
10$. In addition, $M_R=200$ GeV (left) and $M_R=500$ GeV (right).
This calculation was made in a one generation toy model and serves
only for illustration.  The general behaviour is easily understood.
First, recall that within CMSSM roughly $m_{\chi^0_1} \sim m_{\tilde
B} \sim 0.4 M_{1/2}$. Entries on the diagonals of the sneutrino mass
matrix are of the order $m_{LL}^2 \simeq m_{0}^2 + 0.5 M_{1/2}^2$,
$m_{\nu^c\nu^c}^2 \simeq m_0^2+M_R^2$ and $m_{SS}^2 \simeq
m_0^2+M_R^2$.  If $\sqrt{m_0^2+M_R^2} \lsim 0.4 M_{1/2}$ (one of the
pair of) right sneutrinos is the LSP, see left plot. On the other
hand, for larger values of $m_0$ and or $M_R$, right sneutrinos still
can be the LSP if $B_{M_R} \gsim \sqrt{m_0^2+M_R^2}$, since in this
case a large off-diagonal in the sneutrino mass matrix leads to a
large splitting between the two lightest eigenstates, with the lighter
one becoming very light, see right plot. Since $B_{M_R}$ is naively
expected to be of order $m_{SUSY}^2$, sizeable splitting between the
right sneutrinos is expected and in a random scan over parameters such
sneutrinos emerge as LSP quite often. Note, that a light eigenvalue in
the sneutrino sector can also be made by a large off-diagonal in the
sneutrino mass matrix in the $LR$ and $LS$ entries of the mass matrix.

\begin{figure}[htb]
]
\begin{center}
\fcolorbox{white}{white}{
  \begin{picture}(0,100) (630,-15)
    \SetWidth{1.0}
     \SetColor{Black}
    \Line[dash,dashsize=10,arrow,arrowpos=0.5,arrowlength=5,arrowwidth=2,arrowinset=0.2](432,70)(492,30)
    \Line[dash,dashsize=10,arrow,arrowpos=0.5,arrowlength=5,arrowwidth=2,arrowinset=0.2](432,-10)(492,30)
    \Photon(492,30)(552,70){7.5}{4}
    \Photon(492,30)(552,-10){7.5}{4}
    \Line[dash,dashsize=10,arrow,arrowpos=0.5,arrowlength=5,arrowwidth=2,arrowinset=0.2](632,70)(692,70)
    \Line[dash,dashsize=10,arrow,arrowpos=0.5,arrowlength=5,arrowwidth=2,arrowinset=0.2](632,-10)(692,-10)
    \Line[arrow,arrowpos=0.5,arrowlength=5,arrowwidth=2,arrowinset=0.2](692,70)(692,-10)
    \Line[arrow,arrowpos=0.5,arrowlength=5,arrowwidth=2,arrowinset=0.2](692,70)(752,70)
    \Line[arrow,arrowpos=0.5,arrowlength=5,arrowwidth=2,arrowinset=0.2](692,-10)(752,-10)
     \SetWidth{0.5}
     \Text(456,65)[lb]{\Large{\Black{$\tilde{\nu}_{LSP}$}}}
    \Text(434,17)[lb]{\Large{\Black{$\tilde{\nu}_{LSP}$}}}
    \Text(545,43)[lb]{\Large{\Black{$W^+,Z^0$}}}
    \Text(538,10)[lb]{\Large{\Black{$W^-,Z^0$}}}
    \Text(632,75)[lb]{\Large{\Black{${\tilde{\nu}}_{LSP}$}}}
    \Text(632,5)[lb]{\Large{\Black{${\tilde{\nu}}_{LSP}$}}}
    \Text(672,34)[lb]{\Large{\Black{${\tilde{\chi}}^0$}}}
    \Text(722,75)[lb]{\Large{\Black{$f$}}}
    \Text(722,5)[lb]{\Large{\Black{$\bar{f}$}}
         }
    \end{picture}
}
\end{center}
\caption{\label{plot:Feynm2}Examples of Feynman diagrams contributing
to the $\sn\sn$ annihilation: To the left quartic interaction with
gauge bosons; to the right t-channel neutralino exchange.}
\end{figure}
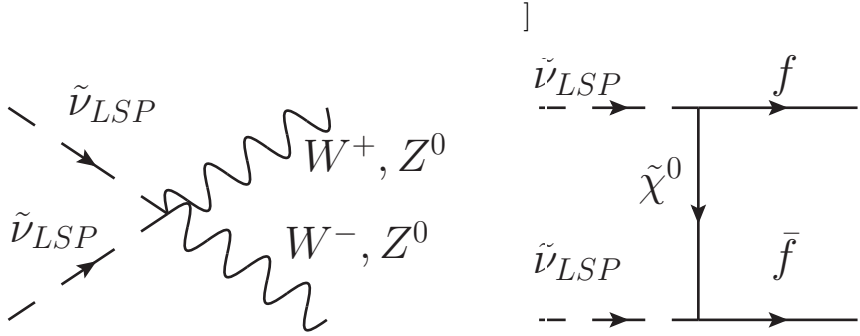

In the early universe sneutrinos can annihilate into SM particles
through various types of interactions. The most important Feynman
diagrams are shown in figs (\ref{plot:Feynm}) and
(\ref{plot:Feynm2}). Fig. (\ref{plot:Feynm}) shows the quartic
interaction between two sneutrinos and two Higgses and s-channel Higgs
exchange. The former is very efficient for $m_{\tilde\nu_{LSP}} \ge
m_{h^0}$, while the latter is important near $m_{\tilde\nu_{LSP}}
\simeq m_{h^0}/2$. Fig. (\ref{plot:Feynm2}) shows the quartic
interaction with W- and Z-bosons and t-channel neutralino
exchange. The importance of the latter depends on the SUSY spectrum.

\begin{figure}[htb]
\begin{center}
\fcolorbox{white}{white}{
  \begin{picture}(361,108) (431,-22)
    \SetWidth{0.5}
    \SetColor{Black}
    \Text(456,65)[lb]{\Large{\Black{$\tilde{\nu}_{LSP}$}}}
    \Text(434,17)[lb]{\Large{\Black{$\tilde{\nu}_{LSP}$}}}
    \Text(515,49)[lb]{\Large{\Black{$h^0$}}}
    \Text(515,12)[lb]{\Large{\Black{$h^0$}}}
    \Text(622,62)[lb]{\Large{\Black{$\tilde{\nu}_{LSP}$}}}
    \Text(618,-14)[lb]{\Large{\Black{$\tilde{\nu}_{LSP}$}}}
    \Text(677,39)[lb]{\Large{\Black{$h^0$}}}
    \Text(755,46)[lb]{\Large{\Black{$f$}}}
    \Text(757,0)[lb]{\Large{\Black{$\bar{f}$}}}
    \SetWidth{1.0}
    \Line[dash,dashsize=10,arrow,arrowpos=0.5,arrowlength=5,arrowwidth=2,arrowinset=0.2](432,76)(481,29)
    \Line[dash,dashsize=10,arrow,arrowpos=0.5,arrowlength=5,arrowwidth=2,arrowinset=0.2](481,29)(533,-22)
    \Line[dash,dashsize=10,arrow,arrowpos=0.5,arrowlength=5,arrowwidth=2,arrowinset=0.2](432,-21)(481,30)
    \Line[dash,dashsize=10,arrow,arrowpos=0.5,arrowlength=5,arrowwidth=2,arrowinset=0.2](481,30)(530,78)
    \Line[dash,dashsize=10,arrow,arrowpos=0.5,arrowlength=5,arrowwidth=2,arrowinset=0.2](593,76)(641,26)
    \Line[dash,dashsize=10,arrow,arrowpos=0.5,arrowlength=5,arrowwidth=2,arrowinset=0.2](593,-21)(640,28)
    \Line[dash,dashsize=10,arrow,arrowpos=0.5,arrowlength=5,arrowwidth=2,arrowinset=0.2](640,28)(726,28)
    \Line[arrow,arrowpos=0.5,arrowlength=5,arrowwidth=2,arrowinset=0.2](726,28)(765,74)
    \Line[arrow,arrowpos=0.5,arrowlength=5,arrowwidth=2,arrowinset=0.2](725,28)(764,-18)
  \end{picture}
}
\end{center}

\caption{\label{plot:Feynm}Examples of Feynman diagrams contributing
to the $\sn\sn$ annihilation: To the left quartic interaction; to the
right s-channel Higgs exchange.}
\end{figure}
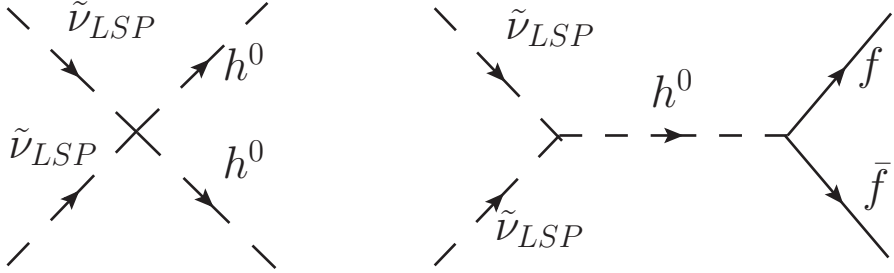

\begin{figure}[htb]
\centering
\includegraphics[scale=0.6]{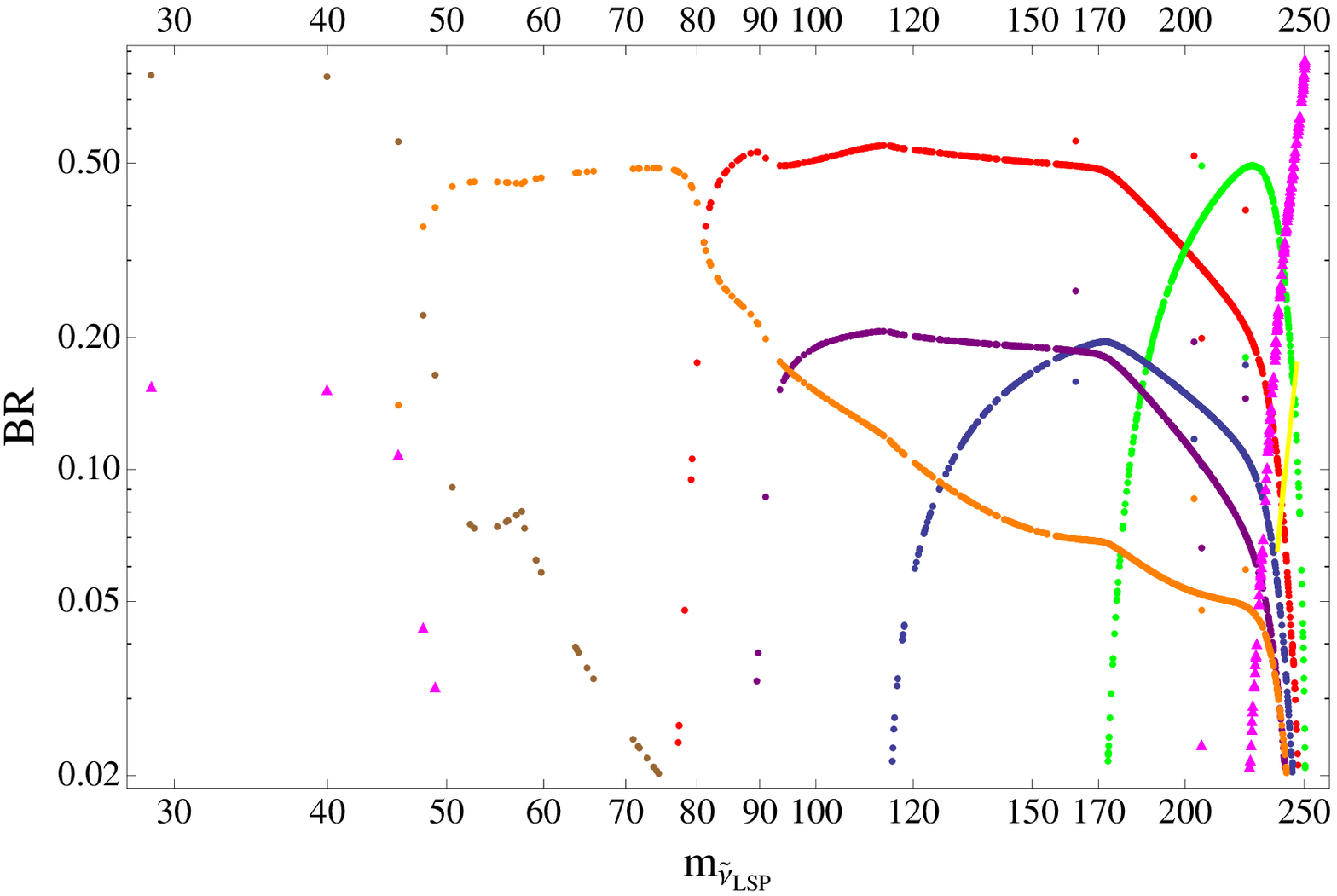}   
\caption{Examples of final state branching ratios for the annihilation 
cross section of sneutrinos to SM final states versus the lightest 
sneutrino mass (in GeV). For the parameter choices of this scan, see text. 
Calculation uses the inverse seesaw model. Different 
kinematical regimes are visible, see discussion. }
\label{fig:BRMR4b}
\end{figure}

The relative importance of different diagrams is strongly dependent on
the kinematical regime.  A typical example of final state branching
ratios versus the lightest sneutrino mass is shown in fig
(\ref{fig:BRMR4b}). In this scan we have fixed $m_0 = 120$,
$M_{1/2} = 600$, $A_0 = 0$ all in GeV and $|Y_{\nu}|=0.4$ and
$\tan\beta = 10$. In addition $\mu_S=[10^{-11},10^{-9}]$ GeV.

From left to right we see that the most important channels are 
$\sn\sn \longrightarrow \tau \overline{\tau}$ (magenta with triangles),
$\sn \sn \longrightarrow b \overline{b}$ (brown), $\sn \sn
\longrightarrow \nu \nu_R$ (orange), $\sn \sn \longrightarrow W^+ W^-$
(red ), $\sn \sn \longrightarrow Z^0 Z^0 $ (purple), $\sn \sn
\longrightarrow H H $ (blue), $\sn \sn \longrightarrow t \overline{t}
$ (green); finally, to the right of the figure, the contributions
coming from the coannihilations are shown: $\tilde{e} \tilde{e}
\longrightarrow \tau \tau$ (magenta with triangles), and $\tilde{e}
\tilde{\overline{e}} \longrightarrow \gamma \gamma$ (in yellow).

For low sneutrino masses the determination of the relic abundance 
is dominated by Higgs exchange, see fig. (\ref{plot:Feynm}) right. 
Since the Higgs couplings are proportional to SM fermion masses, 
$b \overline{b}$ is most important in the low mass regime, followed 
by $\tau \overline{\tau}$. For sneutrino masses above approximately 
$m_\sn \sim 45$ GeV the final state $\nu\nu$ becomes dominant in this 
example. This is because with these parameter choices the lightest 
of the ``singlet'' neutrinos has a mass of about $45$ GeV and the 
Higgs couples always to $\nu_L\nu_R$, i.e. one light and one heavy 
neutrino.

Single $Z^0$ exchange is less important than Higgs exchange, since
scalar-scalar-vector couplings are momentum suppressed. For $m_\sn
\gsim 80$ GeV, however, two gauge boson final states become dominant,
the channel $W^+ W^-$ being more important than $Z^0 Z^0 $. For masses
above $m_\sn \gsim 120$ GeV also two Higgs final states are
sizable. All these final states are due to quartic interactions, see 
fig. (\ref{plot:Feynm}) left and fig. (\ref{plot:Feynm2}). 
Due to the large top Yukawa coupling, the two top final state, 
once kinematically possible, becomes very important. And, finally, 
for $m_\sn$ approaching the NLSP mass, in this example the lightest 
scalar tau, coannihilation into taus becomes dominant.

\begin{figure}
\centering
\includegraphics[scale=0.7]{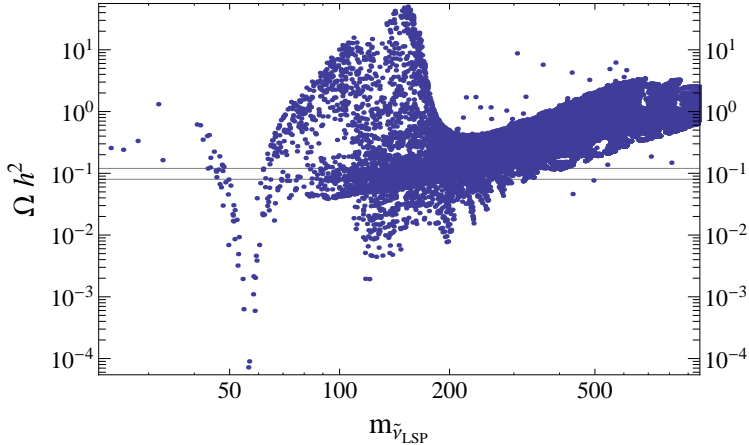}   
\caption{General scan for the inverse seesaw model. The plot shows 
$\Omega h^2$ versus the mass of the lightest sneutrino (in GeV) for points 
in which the sneutrino is the LSP.}
\label{fig:gen}
\end{figure}

Next, we have performed a general scan over the parameter space of the
model choosing randomly ($m_0,M_{1/2},A_0,\tan\beta,{\rm sgn}(\mu)$)
in the interval $m_0 =[100,3000]$, $M_{1/2}=[200,3000]$, $A_0=0$,
$\tan\beta=10$ and ${\rm sgn}(\mu)>0$ and $|Y_{\nu}|=0.3$, $M_R
=[0,1000]$. $B_{M_R}$ is calculated accordingly to enhance the
percentage of sneutrino LSP points. We post-select data points with
sneutrino LSPs and cut on all points not fulfilling the lower bounds
on squark and gluinos masses from the LHC \cite{ATLAS-CONF-2012-109}.
Results are shown in fig. (\ref{fig:gen}) for the case of the inverse
seesaw. Shown is the calculated RA ($\Omega h^2$) versus the mass of
the lightest sneutrino for points in which the sneutrino is the LSP.
The band, which is the allowed range from WMAP \cite{Larson:2010gs}, 
shows that one can
easily get points with the correct relic abundance over a wide range
of parameters. The figure is for the inverse seesaw, linear seesaw is
qualitatively very similar.

The plot shows several distinct features. First, for masses of
sneutrinos around $m_\sn \simeq 60$ GeV a strong reduction of the RA
occurs, due to the s-channel Higgs exchange. As can be seen, this
diagram is very effective in reducing the RA whenever $m_\sn$ is
within a few GeV of the mass of the Higgs, but less important
elsewhere. In the region above $m_\sn = 80$ GeV, quartic interactions
with the gauge bosons are effective and above $m_\sn = 175$ GeV
two-top final states become dominant. For very large $m_\sn$ one sees
an overall trend that the RA rises with rising sneutrino mass, apart
from a few scattered points. Low RA, i.e. $\Omega h^2 \simeq 0.1$, in
this high mass regime can practically only be made via co-annihilation
or s-channel heavy Higgs exchange. Note that the fact that there are
only a few points with $m_\sn$ below 50 GeV is just an artifact of the
scanning procedure. However, the general trend is that for very light
sneutrinos the calculated RA is larger than $\Omega h^2 \sim 0.1$.
We will come back to a more detailed discussion
of light sneutrinos in the next section.

\begin{figure}
\centering
\includegraphics[scale=0.6]{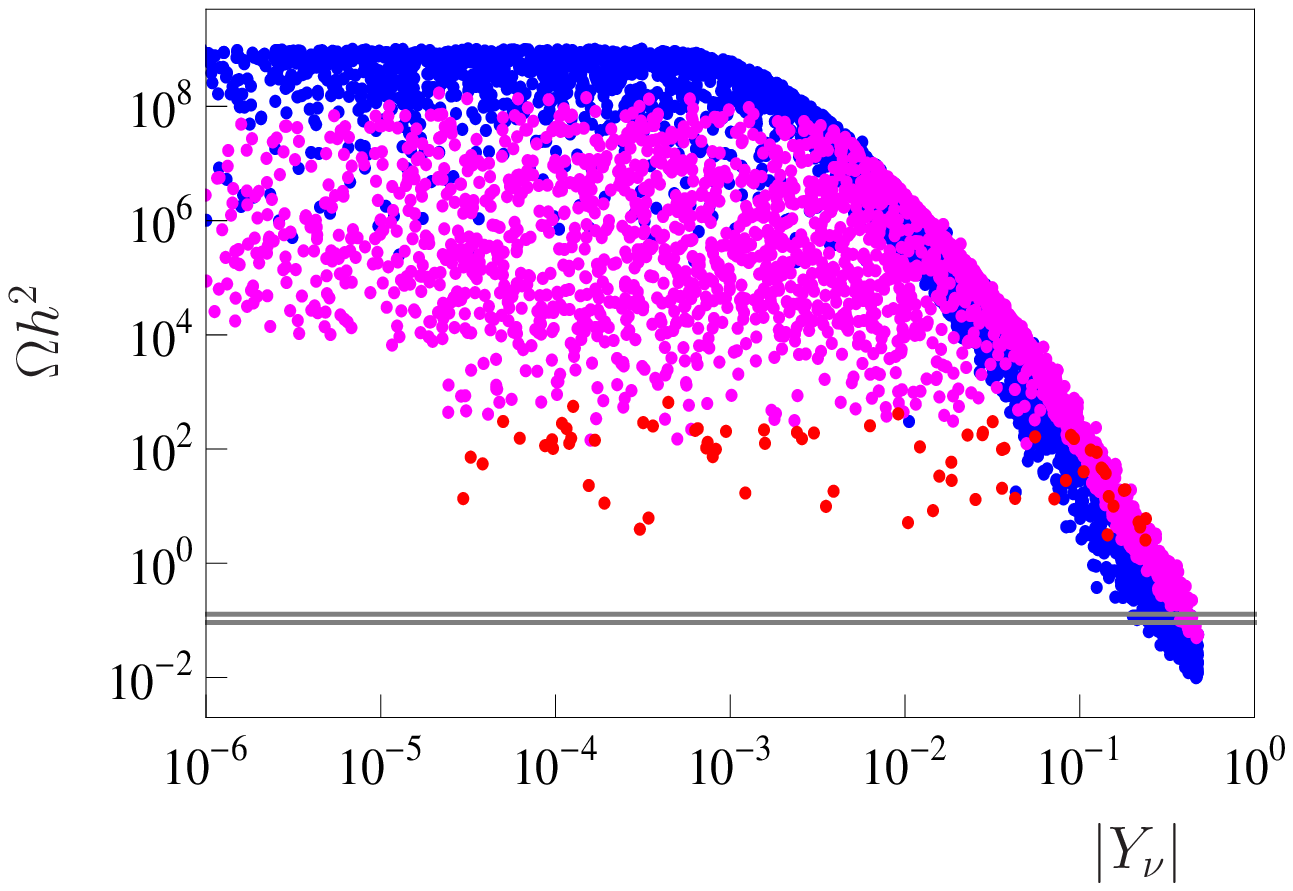}
\includegraphics[scale=0.6]{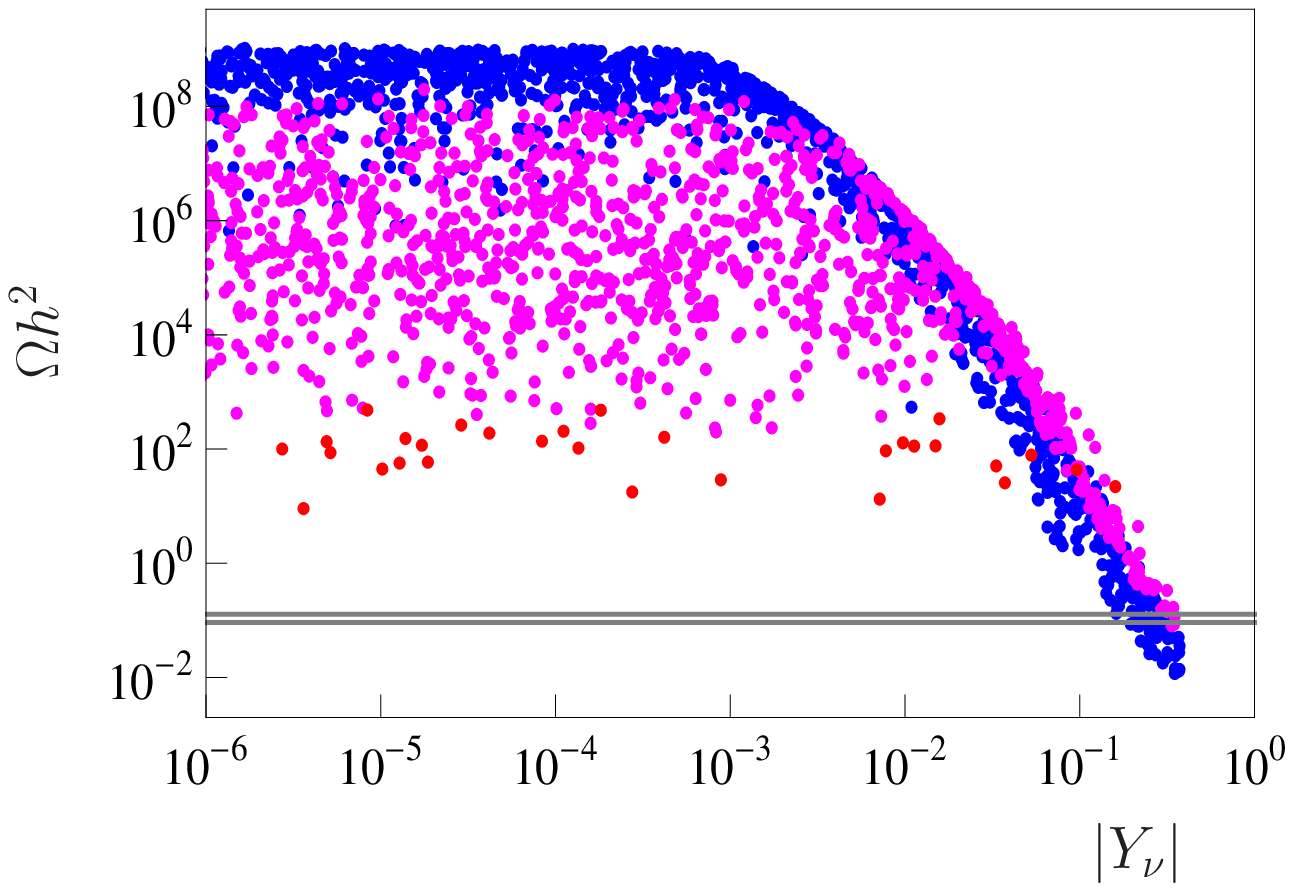}
\caption{Scan for the inverse (left) and linear (right) seesaw model. 
The plot shows $\Omega h^2$ versus $|Y_{\nu}|$ in a scan over the 
remaining parameters, see text. The color coding of the points shows 
the mass difference between the lightest sneutrino mass and the 
NLSP (next-to-LSP) mass, in this scan practically always the lightest 
of the charged sleptons. Red: $m_{NLSP}-m_{LSP}<100$ GeV, violet: 
$100< m_{NLSP}-m_{LSP}<500$ GeV, blue: $m_{NLSP}-m_{LSP}>500$ GeV. 
For a discussion see text.}
\label{fig:Yuk}
\end{figure}

In fig. (\ref{fig:gen}) we have fixed the neutrino Yukawa couplings to
a constant value. However, sneutrinos which are purely singlets do not
couple to gauge bosons and thus their relic abundance is usually too
large. For mixed sneutrinos the RA depends strongly on the choice of
$|Y_{\nu}|$. An example is shown in fig. (\ref{fig:Yuk}). The figure
shows on the left (right) results for the inverse (linear) seesaw.  In
both cases we have fitted neutrino data, using eqs (\ref{eq:ISdirac})
and (\ref{eq:YvuLin}), and scanned over the parameters: $m_0$ and
$M_{R}$ in the interval [100,1000] GeV and $B_{M_R}=[10^3,10^6]$
GeV$^2$. Here, $M_{1/2}$ was fixed to $M_{1/2}=2.5$ TeV and
$\tan\beta=10$ and $A_0=0$.  The choice of such a large $M_{1/2}$
guarantees that all points have a lightest Higgs mass in the vicinity
of 125 GeV. It also makes all SUSY particles, except the sneutrino,
relatively heavy.

The points in fig. (\ref{fig:Yuk}) are color coded by the mass
difference between the lightest sneutrino mass and the NLSP
(next-to-LSP) mass, in this scan practically always the lightest of
the charged sleptons. Red: $m_{NLSP}-m_{LSP}<100$ GeV, violet: $100<
m_{NLSP}-m_{LSP}<500$ GeV, blue: $m_{NLSP}-m_{LSP}>500$ GeV.  For
large $|Y_{\nu}|$ the RA goes down as $\Omega h^2 \propto
|Y_{\nu}|^{-4}$, for small values of $Y_{\nu}$ the points show
practically no dependence on $|Y_{\nu}|$. This is because the
determination of the RA is then dominated by coannihilation processes
with the lightest stau.  These can be very efficient, if $\Delta m^2 =
{m_{\tilde{l}}}^2 - {m_{\sn}}^2 \simeq$ few GeV, less so for larger
mass differences. Thus, to reduce the relic density of the sneutrino
to acceptably small values, one needs either a special kinematic
configuration, such as co-annihilation or s-channel resonance, or
$|Y_{\nu}|$ has to be larger than roughly $|Y_{\nu}|\gsim 0.1$.

\subsubsection{Direct Detection}

Direct detection of the sneutrinos consists in detecting the recoil
energy coming from the elastic scattering of sneutrinos with nuclei
inside a detector. The interaction, which occurs in the non
relativistic limit, since the velocity of dark matter
particles in the Galactic halo is small, comes from basically two
diagrams contributing at tree level: the t-channel exchange
of a neutral Higgs or of the Z boson. Which of the two diagrams is 
the more important one depends on the actual value of $|Y_{\nu}|$. 

\begin{figure}
\begin{center}
\fcolorbox{white}{white}{
  \begin{picture}(169,170) (46,-23)
    \SetWidth{1.0}
    \SetColor{Black}
    \Line[dash,dashsize=10,arrow,arrowpos=0.5,arrowlength=5,arrowwidth=2,arrowinset=0.2](47,146)(117,97)
    \Line[dash,dashsize=10,arrow,arrowpos=0.5,arrowlength=5,arrowwidth=2,arrowinset=0.2](117,97)(193,145)
    \Line[dash,dashsize=10,arrow,arrowpos=0.5,arrowlength=5,arrowwidth=2,arrowinset=0.2](119,96)(119,25)
    \Line[arrow,arrowpos=0.5,arrowlength=5,arrowwidth=2,arrowinset=0.2](51,-21)(119,26)
    \Line[arrow,arrowpos=0.5,arrowlength=5,arrowwidth=2,arrowinset=0.2](119,26)(189,-22)
    \Text(47,104)[lb]{\Large{\Black{${\tilde{\nu}}_{LSP}$}}}
    \Text(180,106)[lb]{\Large{\Black{${\tilde{\nu}}_{LSP}$}}}
    \Text(130,65)[lb]{\Large{\Black{$Z^0, H^0$}}}
    \Text(81,-16)[lb]{\Large{\Black{$q$}}}
    \Text(177,-2)[lb]{\Large{\Black{$\bar{q}$}}}
  \end{picture}
}
\end{center}
\caption{Diagrams contributing to the direct detection cross section:
elastic scattering of $\sn$ over quarks.}
\label{fig:BR}
\end{figure}
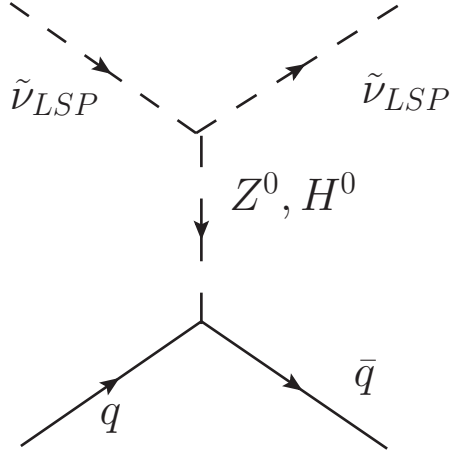

The $Z$--boson exchange cross section is \cite{Arina:2007tm}:
\begin{equation}
\label{eq:DDZ}
 \sigma_{\sn \; \cal N}^{Z}=\frac{G_F^2}{2\pi}\frac{m^2_{\sn}m^2_{\cal N}}{(m_\sn+m_{\cal N})^2}
 \left[ A_{\cal N}+2(2\sin^2\theta_W-1)Z_{\cal N}\right]^2
\end{equation}
where $m_{\cal N}$ is the nucleus mass, $A_{\cal N}$ and $Z_{\cal N}$
are the mass number and proton number of the nucleus and $G_{F}$ is
the Fermi constant.

The Higgs--bosons exchange scattering cross section is \cite{Arina:2007tm}:
\begin{equation}
 \sigma_{\sn \; \cal N}^{Higgs}= \frac{m^2_{p}}{4\pi (m_{\sn}+m_{\cal N})^2} 
\left[ f_p Z_{\cal N}+f_n\left(A_{\cal N}-Z_{\cal N}\right) \right]^2
 \label{eq:DDhiggs}
\end{equation}
where $\cal N$ denotes the nucleus, and the quantities $A_{\cal N}$
and $Z_{\cal N}$ are the mass number and proton number of the nucleus,
$f_{p}$ and $f_{n}$ are hadronic matrix elements which parametrize
the quark composition of the proton and the neutron, and which
represent the effective coupling of the $\sn$ to the nucleus, but are
subject to considerable uncertanties 
\cite{Arina:2007tm,Bottino:2001dj}.

\begin{figure}[htb]
\centering
\includegraphics[scale=0.7]{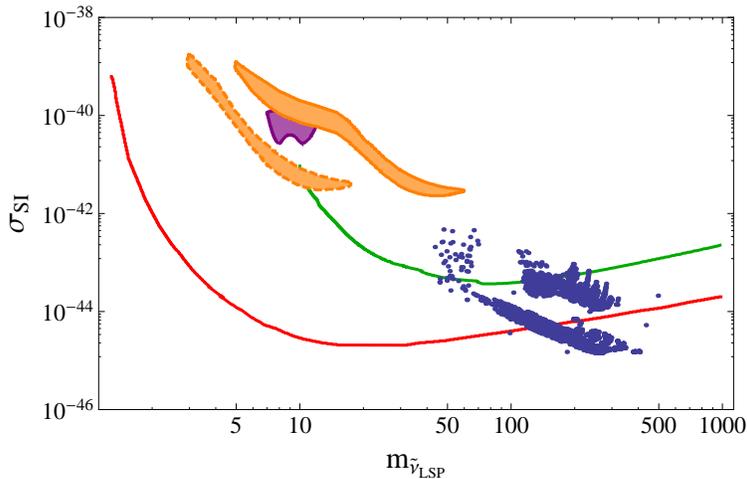}   
\caption{\label{fig:genDD} Direct detection cross section (in
[$cm^2$]) for sneutrino LSPs (masses in GeV), for the inverse seesaw
model. The points are those from fig. (\ref{fig:gen}) compatible with
the upper bound on the relic abundance. Also the current limits from
XENON-100 \cite{Aprile:2012vw} (red line), CDMS \cite{Ahmed:2009zw}
(green line), DAMA (with and without channeling, orange regions)
\cite{Bernabei:2010mq}, and Cogent \cite{Aalseth:2010vx} (purple
region) are shown for comparison.}
\end{figure}

In figure (\ref{fig:genDD}) we depict the direct detection cross section
versus the LSP sneutrino mass (blue points). The points are the same
as shown in fig. (\ref{fig:gen}), but after a cut on the relic
abundance. In the same plot, the current limits from XENON-100
\cite{Aprile:2012vw} (red line), CDMS \cite{Ahmed:2009zw} (green
line), DAMA (with and without channeling, orange regions)
\cite{Bernabei:2010mq}, and Cogent \cite{Aalseth:2010vx} (purple
region) are shown.  The major bound nowadays comes from the XENON-100
experiment \cite{Aprile:2012vw}, whose best sensitivity is around
$10^{-44} \rm cm^2$ for a dark matter candidate of 50 GeV. The
sneutrinos show a SI cross section $\sigma_{SI} \lesssim 10^{-42} \rm
cm^2$, and for masses $m_{\sn} \gtrsim 100$ GeV they are compatible
with current limits by XENON-100. However, XENON-1T, whose sensibility
should improve up to $10^{-46} \rm cm^2$, will test those cross
sections.

We have not been able to find low sneutrino masses of the
order of ${\cal O}(5-10)$ GeV, which have the correct relic density
and fulfill at the same time the constraints from the direct
detection experiment XENON-100 \cite{Aprile:2012vw}. However, this
calculation has been done with $B_{\mu_S} \propto m_0\mu_S$ and 
lepton number violation in the sneutrino mass matrix leads to the mass
splitting between the real and the imaginary part of the lightest
sneutrino, and the scattering via Z boson exchange occurs
inelastically, through a transition from the real to the imaginary or
viceversa. Points shown in fig. (\ref{fig:genDD}) have all very small
splitting in the sneutrino sector, but if the mass splitting is
greater than some keV, scattering is strongly suppressed at direct
detection experiments. Indeed, the maximum kinetic energy that the
sneutrino LSP can transfer to the detector depends on the velocity it
moves relative to the nucleus $v$ ($\simeq 10^{-3}$ in the galactic
halo),the nucleus mass M and the angle $\theta$ of scattering:

\begin{equation}
E = \frac{A^2 v^2}{M} (1-cos(\theta))
\end{equation} 

where $A = \frac{m_{\sn} M}{m_{\sn} + M}$, which would give, in the
case of a Xenon detector for instance, and $m_{\sn} = 100$ GeV, E = 25
keV (if cos($\theta$) = 0). For heavier sneutrinos with a mass of the
order of TeV, for a splitting larger than some hundred keV the direct 
detection cross section goes to zero. Such ``large'' splitting is 
currently not excluded in the inverse seesaw, compare to fig. 
(\ref{fig:split}). Thus, in principle inverse seesaw can evade all 
constraints from direct detection, while linear seesaw can not, see the 
discussion in section (\ref{subesct:linss}).

\subsection{mBLR model}
\label{sec:snudmBLR}

In this subsection we discuss the DM phenomenology of the
supersymmetric $U(1)_R \times U(1)_{B-L}$ extension of the standard
model. The main difference to the simpler models discussed previously
are the presence of the extra gauge boson $Z'$ and the possibility to
have an additional light, mostly singlet Higgs boson, which lead to
some important changes in the phenomenology.

First, recall that the $U(1)_{R}\times U(1)_{B-L}$ gauge symmetry of
this model is spontaneously broken to the hypercharge group $U(1)_{Y}$
by the vevs $v_{\chi_{R}}$ and $v_{\bar\chi_{R}}$ of the scalar
components of the $\hat\chi_R$ and $\hat{\bar{\chi}}_R$ superfields
whereas the $SU(2)_{L}\otimes U(1)_{Y}\to U(1)_{Q}$ breaking is driven
by the vevs $v_{d}$ and $v_{u}$ of the neutral scalar components of
the $SU(2)_L$ Higgs doublets $H_d$ and $H_u$ up to gauge kinetic
mixing effects. The tadpole equations for the different vevs can be
solved analytically for either (i) ($\mu,B_\mu, \mu_R$, $B_{\mu_R}$)
or (ii) ($\mu, B_\mu$, $m^2_{\chi_R}$,$m^2_{{\bar\chi}_R}$) or (iii)
($m^2_{H_d}$, $m^2_{H_u}$, $m^2_{\chi_R}$, $m^2_{{\bar\chi}_R}$)
\cite{Hirsch:2012kv}.

We address the minimal version option (i) as
CmBLR (constrained mBLR), since it allows to define boundary
conditions for all scalar soft masses at $m_{GUT}$, reducing the
number of free parameters by four, although leading to some
constraints on the parameter space, such as a lower bound on
tan$\beta_R$ (tan$\beta_R >1$) \cite{Hirsch:2012kv}. The second option
(ii) is instead more flexible, and we have made use of it in some of
our scans, too. We will refer to this option as $\chi_R$mBLR
version (non-universal $\chi_R$ masses mBLR). We have not used the last
option, which we only mentioned for the sake of completeness.

\begin{figure}[t]
\centering
{\includegraphics[scale=0.55]{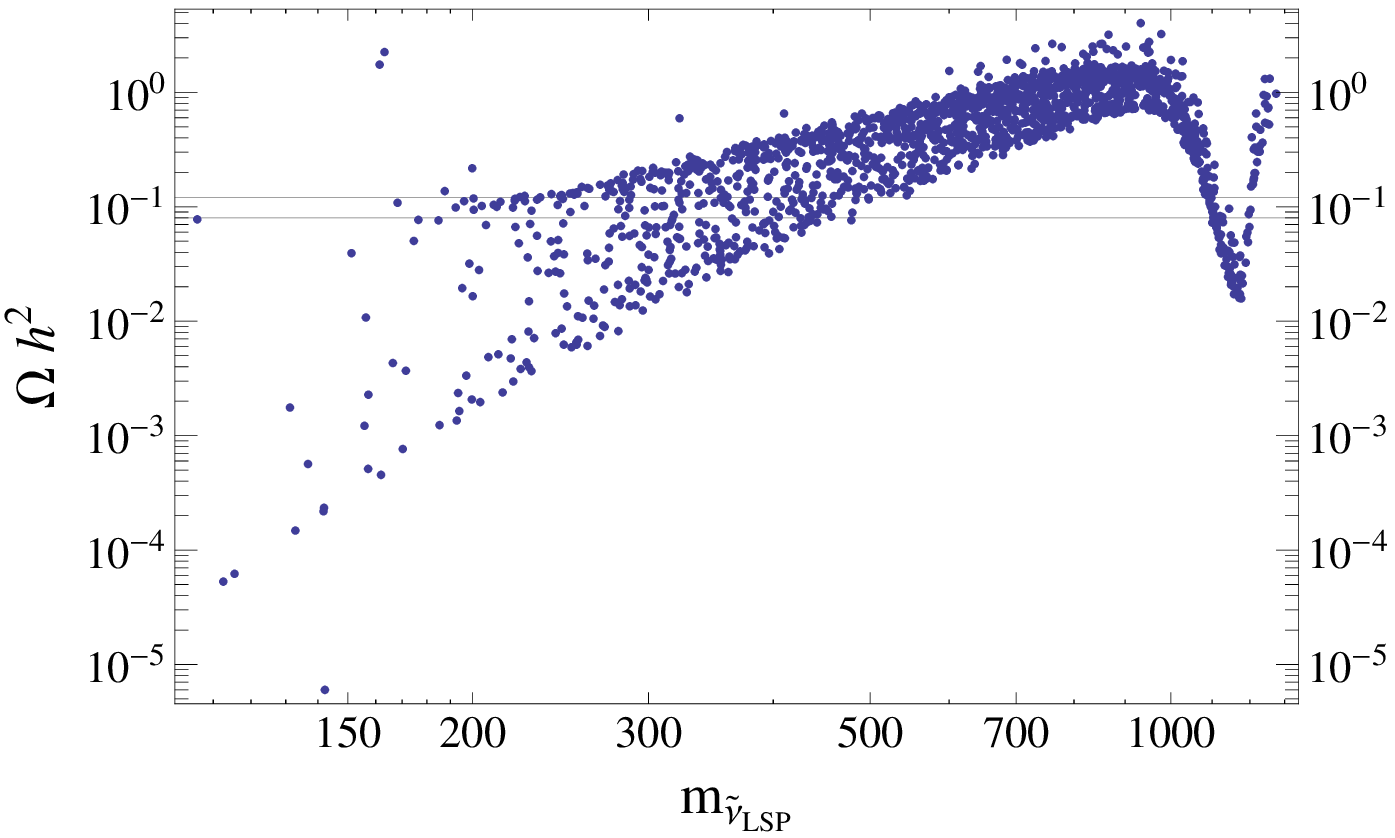}}
{\includegraphics[scale=0.55]{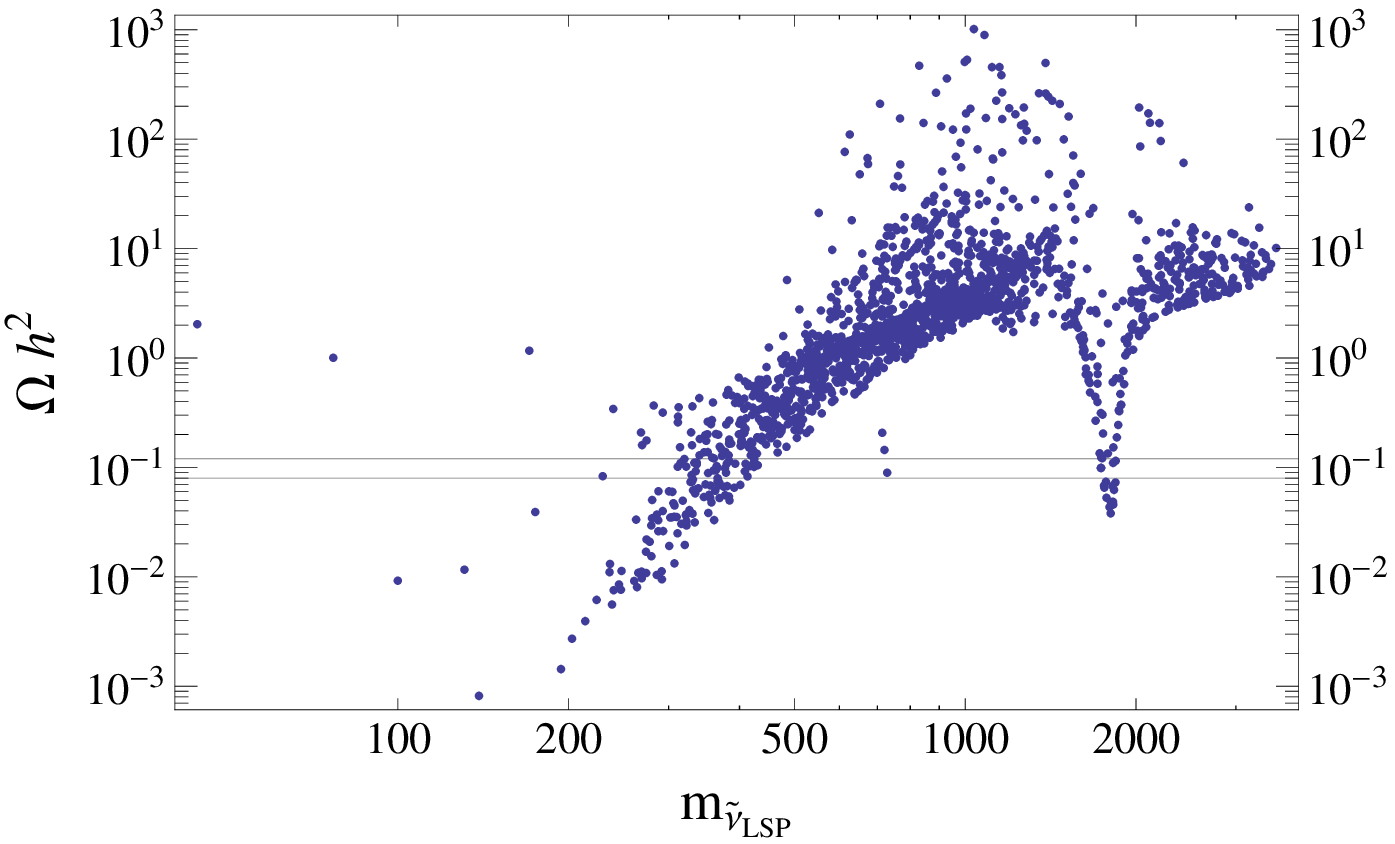}}
\caption{\label{fig:BLRDMgenvR}Scan for the CmBLR version of the
extended gauge model. Parameters are varied as follows: $m_0 =
[0,6000]$ GeV, $M_{1/2} = [3000,8000]$ GeV, tan$\beta_R$ =
[1.0,1.3]. The other parameters are set to the values tan$\beta$=10,
$A_0 = -4500$ GeV, $Y_S =$ diag(0.3); $v_R$ has been chosen different
in the two plots, $v_R = 6$ TeV and $v_R = 10$ TeV,
respectively. Masses of the $\sn$ are in GeV. }

\end{figure}

The result of $\Omega h^2$ for two general scans is shown in
fig. (\ref{fig:BLRDMgenvR}). Parameters have been scanned as described
in the figure caption. Note that there are two fixed but different
choices of $v_R$ in the left and right plots, leading to two different
values of the $Z'$ mass. In both plots in fig. (\ref{fig:BLRDMgenvR})
the main feature clearly visible is the $Z'$ pole. Indeed, the
annihilation of the $\sn$ LSPs into SM particles via the $Z'$ becomes
efficient when the mass of the $\sn$ is close to half the mass of the
$Z'$. The mass of the $Z'$ can be calculated analytically
\cite{Hirsch:2012kv} and mainly depends on the value of $v_R$, see
eq.(\ref{eq:mZp}). The ATLAS searches for a $Z'$ set a lower limit on
its mass which is 1.8 TeV, and this translates into a lower limit on
$v_R \gtrsim 5$ TeV, see the plot on the left. The plot on the right
shows that choosing a higher value of $v_R$ we can get very heavy
$\sn$ DM with the correct RA, up to masses of several TeV.

\begin{figure}[t]
\centering
{\includegraphics[scale=0.41]{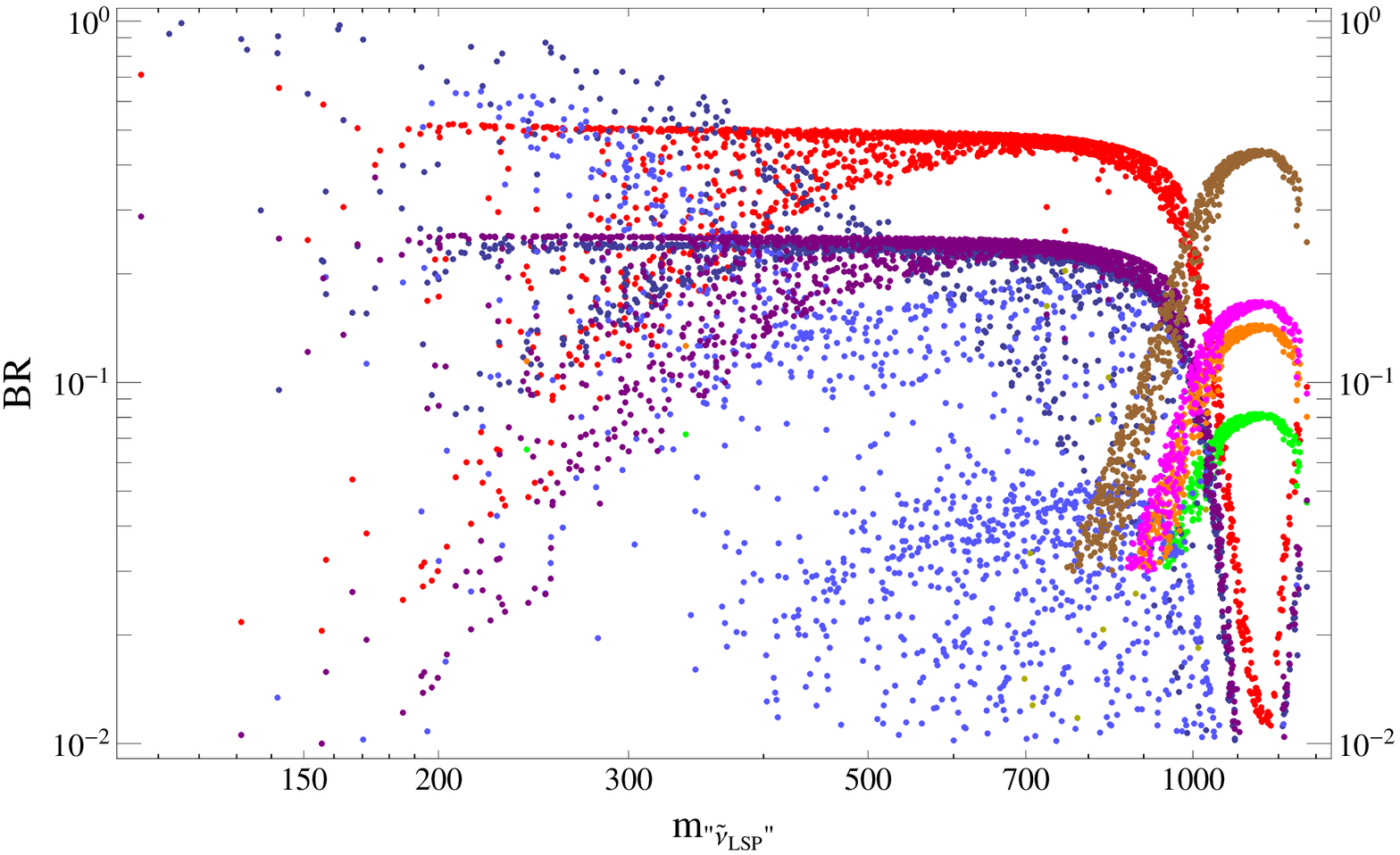}}
{\includegraphics[scale=0.41]{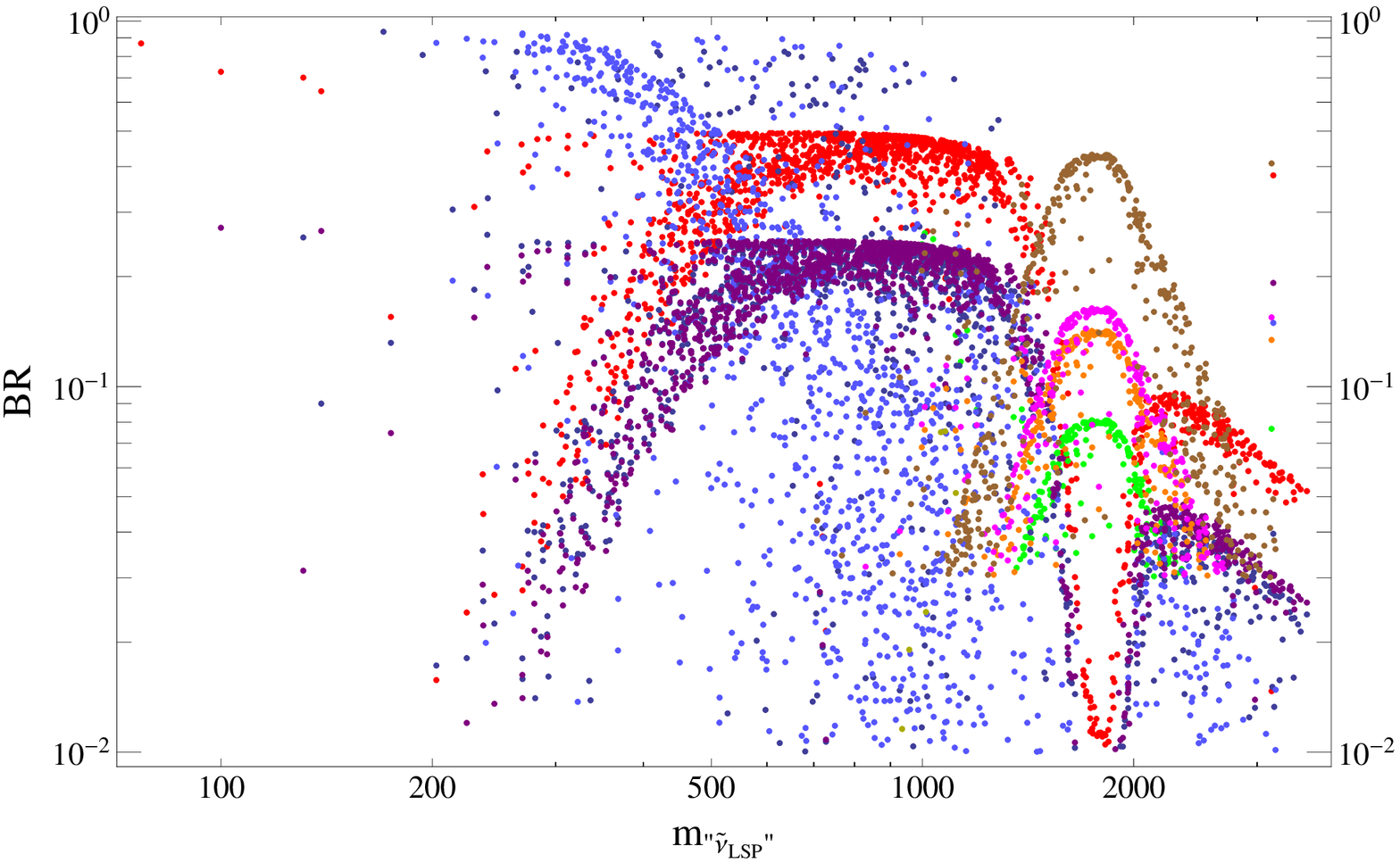}}
\caption{\label{fig:BRBLRDMgenvR} Final state branching ratios
for the annihilation cross section of sneutrinos to SM final states
versus the lightest sneutrino mass (in GeV). For the parameter choices
of these scans see fig. (\ref{fig:BLRDMgenvR}). For a discussion of the
different kinematical regimes which are visible, see text.}
\end{figure}

The main annihilation channels for sneutrino DM for the points of
fig. (\ref{fig:BLRDMgenvR}) are shown in
fig. (\ref{fig:BRBLRDMgenvR}). Far from the $Z'$-pole resonance these
are $\sn \sn \longrightarrow \tau \overline{\tau}$ (magenta ), $\sn
\sn \longrightarrow b \overline{b}$ (brown), $\sn \sn \longrightarrow
\nu \nu_R$ (orange), $\sn \sn \longrightarrow W^+ W^-$ (red), $\sn \sn
\longrightarrow Z^0 Z^0/Z_R Z_R $ (purple ), $\sn \sn \longrightarrow
H H $ (blue), $\sn \sn \longrightarrow t \overline{t} $ (green).  The
quartic coupling with two Higgses ( $h^0$, $h^0_{BLR}$ and $A^0$,
depending on if they are kinematically allowed, depending on the $\sn$
mass) is one of the most efficients, as before. For lower masses the
annihilation via the MSSM Higgs is the most efficient, as can be
noticed by the small relic density for lower masses,
expecially in the first plot, where on the left end side we are
approaching the region where the quartic Higgs coupling is
important (for $m_{\sn} \simeq 120$ GeV).

Recall that in this model the Higgs sector is more complicated due to the
extended gauge structure. The $U(1)_{B-L}\times U(1)_{R}$ breaking
results in one additional light Higgs, $h^0_{BLR}$ 
\cite{Hirsch:2011hg}. The mixing between the MSSM Higgs $h^0$ and the
$h^0_{BLR}$ enhances the mass of the mostly MSSM Higgs, leading to 
a MSSM-like Higgs in accord with the most recent ATLAS and CMS 
preferred regions, without much constraints on the SUSY spectrum. 
However, this enhancement of the MSSM Higgs mass occurs usually in 
the model if the $h^0_{BLR}$ has a mass of the order of the MSSM-like 
state or less, i.e. the presence of a light singlet Higgs is preferred 
unless the SUSY spectrum is rather heavy (in which case the CMSSM 
limit is reached).

\begin{figure}
\centering
\includegraphics[scale=0.7]{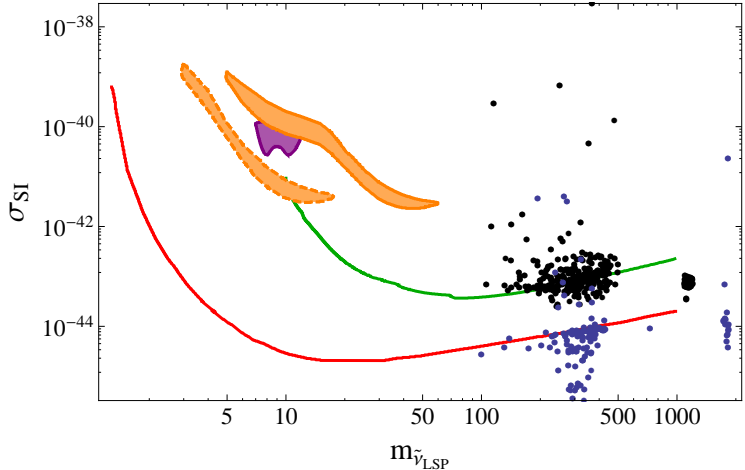}  
\caption{\label{fig:BLRDD}  Direct detection cross section (in
$cm^2$) for sneutrino LSPs in the BLR model. Masses are in GeV. Black
points refer to the scan described in fig.(\ref{fig:BLRDMgenvR}) left
with $v_R=6$ TeV. Blue points stand for the scan of
fig.(\ref{fig:BLRDMgenvR}) right with $v_R=10$ TeV.  All points shown
fulfill the RA constraints. Higher $v_R$ leads in general to lower DD
cross section.}
\end{figure}

We have also checked for constraints coming from direct detection 
in the limit of negligible sneutrino splitting. Examples for 
direct detection cross section are shown in fig. (\ref{fig:BLRDD}). 
As before, see fig.(\ref{fig:genDD}), different experimental constraint 
are also shown. All points shown fulfill the constraints from relic  
abundance. We have calculated two scans, one with $v_R=6$ TeV (black) 
and one with  $v_R=10$ TeV (blue). As can be seen, practically all of 
the points  with $v_R=6$ TeV in this scan are excluded by the limit 
from XENON-100, while most of the $v_R=10$ TeV are allowed. Thus 
XENON-100 puts currently a lower bound on $v_R$ (and thus the $Z'$ mass) 
of the order of  $v_R \simeq 10$ TeV for sneutrino LSPs as DM. 

The origin of this surprisingly strong constraints lies in the 
$Z^0-Z'$ mixing. The mixing angle between these two states is 
roughly of the order $\theta_{Z^0Z'} \sim (g_Lv^2)/(g_Rv_R^2)$. 
Thus the right sneutrinos, which couple mostly to the $Z'$, couple 
via this mixing also to the $Z^0$. The $Z^0$ has an experimentally 
fixed mass. Thus, the only possibility to suppress the DD cross 
section \footnote{Apart from a large sneutrino splitting.} is to 
increase $v_R$.

\begin{figure}[htb]
\centering
\includegraphics[scale=0.7]{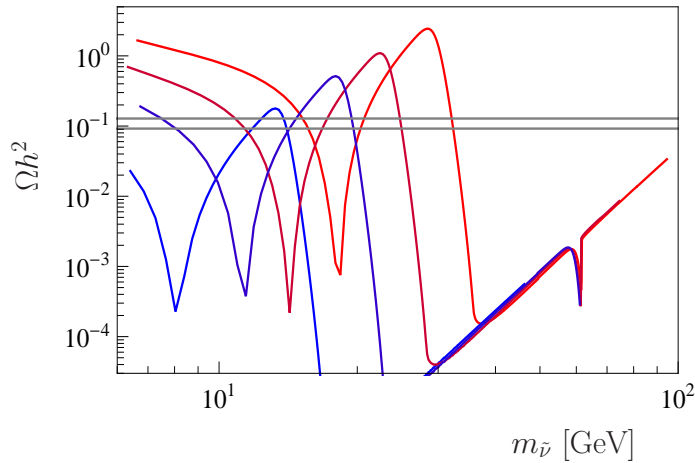}
\caption{Scan into the low sneutrino mass region using the mBLR model. 
For the parameter choices see text.  The figure demonstrates that 
the mBLR mode can give the correct RA for low mass sneutrinos in 
those parts of the parameter space where a light, singlet Higgs 
is present. }
\label{fig:LowSnu}
\end{figure}

Finally, fig. (\ref{fig:LowSnu}) shows a dedicated scan for low mass
sneutrinos in the mBLR model. The different curves are slight
variations of the parameters near the study point BLRSP1. The original
parameters of BLRSP1 were: $m_0=470$, $M_{1/2}=700$, $\tan\beta=20$,
$A_0=0$, $v_R=4700$, $\tan\beta_R=1.05$, $\mu_R=-1650$ and
$M_{A_R}=4800$ GeV.  To obtain very light sneutrinos, $m_0$ has been
lowered to $m_0=440 $ GeV, while the different curves are for
$M_{1/2}=650,660,675$ and $700$ GeV and the scanning runs $m_{A_R}$
from $3000-4000$ GeV. The resulting scan produces sneutrinos with
masses in the interval [$5,100$] GeV, while the lightest Higgs mass,
in this case a mostly singlet Higgs, has a mass eigenvalues of
$m_{h_1} \simeq 1-50$ GeV. The figure shows a pole around $m_\sn
\simeq 62$ GeV, due to a mass for the MSSM-like Higgs of around
$124-125$ GeV in all cases. There appear additional dips in the RA for
smaller sneutrino masses, whenever $m_\sn \simeq m_{h_1}/2$. This
demonstrates that in the extended gauge model it is possible to have
the correct RA even for very low sneutrino masses.

However, note that, while the model can in principle give DD
cross section large enough to explain the DAMA \cite{Bernabei:2010mq},
and Cogent \cite{Aalseth:2010vx} hints, such points will 
always be inconsistent with the constraints from XENON-100 
\cite{Aprile:2012vw}, also for the case of inelastic dark 
matter \cite{Aprile:2011ts}.

\section{Conclusions}
\label{sec:cncl}

We have studied low scale seesaw models with a sneutrino LSP. We 
considered two possibilities: Models with the MSSM gauge group 
and either a linear or inverse seesaw and a model with the 
gauge group  $SU(3)_c \times SU(2)_L \times U(1)_{B-L}\times U(1)_R$ 
and an inverse seesaw. Sneutrinos can be the DM in both cases, 
fulfilling all known experimental bounds.

However, while inverse and linear seesaw lead to different results for
LFV, in general, they give similar DM results. 
There are some differences in detail, though: 
In the inverse seesaw it is possible to avoid all direct detection 
constraints using a large enough splitting in the sneutrino 
sector, which leads to ``inelastic'' dark matter. This is not 
possible in the linear seesaw, due to constraints from neutrino 
physics. 

In the extended gauge model there is more freedom than in the 
simpler MSSM-group based models. Especially very light (${\cal O}(1)$ 
GeV) or very heavy (${\cal O}({\rm several})$ TeV) sneutrinos can 
give the  correct relic density, due to the existence of a 
mostly singlet Higgs in the former case and due to the $Z'$ 
in the latter. Very light sneutrinos could explain the hints 
from DAMA \cite{Bernabei:2010mq} or Cogent \cite{Aalseth:2010vx}, 
but are inconsistent then with  XENON-100 \cite{Aprile:2012vw,Aprile:2011ts}.

Finally, it is interesting to note that in the limit of small sneutrino mass
splitting the DD limit from XENON-100 \cite{Aprile:2012vw} leads to a
lower limit on $v_R$ of the order of ${\cal O}(10)$ TeV for sneutrino
LSPs as the dominant component of the galactic dark matter.

\section*{Acknowledgements}

We thank Werner Porod and Florian Staub for helpful discussions and
assistance with SPheno and SARAH. We acknowledge support from the Spanish
MICINN grants FPA2011-22975, MULTIDARK CSD2009-00064 and by the
Generalitat Valenciana grant Prometeo/2009/091 and the EU~Network
grant UNILHC PITN-GA-2009-237920.



\end{document}